\documentclass[journal]{IEEEtran}

\usepackage{graphicx} \usepackage{listings}
\usepackage[ruled,vlined,linesnumbered,noend]{algorithm2e}
\usepackage[inline]{enumitem}
\usepackage{xspace}
\usepackage{xcolor}
\usepackage{todonotes}
\usepackage{multirow}
\usepackage[normalem]{ulem}
\useunder{\uline}{\ul}{}
\usepackage{subcaption}

\newcommand{\iotatrace}{$\iota$-trace\xspace}
\newcommand{\fslice}{f-slice\xspace}
\usepackage{hyperref}

\usepackage[T1]{fontenc}
\usepackage[scaled=0.85]{beramono}
\usepackage{siunitx}
\usepackage{amsmath}
\usepackage{xfrac}    \usepackage{amsthm}
\theoremstyle{definition}
\newtheorem{definition}{Definition}[section]

\usepackage{array,graphicx}

\usepackage[most]{tcolorbox}

\definecolor{mainColor}{rgb}{0.16, 0.67, 0.53}
\definecolor{minorColor}{rgb}{0.6, 0.4, 0.8}

\newcommand{\icftl}{iCFTL\xspace}
\newcommand{\icftlDiagnostics}{\texttt{iCFTLdiagnostics}\xspace}

\newcommand{\precision}[0]{p}
\newcommand{\recall}[0]{r}

\renewcommand{\paragraph}[1]{{\bf #1:}}
\usepackage{amssymb}
\usepackage{mathrsfs}
\usepackage[numbers]{natbib}

\usepackage{fancybox} 
\begin{document}
    \title{Diagnosing Violations of State-based Specifications in iCFTL}

\author{Cristina Stratan, Claudio Mandrioli, Domenico Bianculli

\thanks{C. Stratan, C. Mandrioli and D. Bianculli are with the University of Luxembourg, Luxembourg}}

     \maketitle
    \begin{abstract}
As modern software systems grow in complexity and operate in dynamic environments, the need for runtime analysis techniques becomes a more critical part of the verification and validation process.
    Runtime verification monitors the runtime system behaviour by checking whether an execution trace---a sequence of recorded events---satisfies a given specification, yielding a Boolean or quantitative verdict.
However, when a specification is violated, such a verdict is often insufficient to understand why the violation happened.
    To fill this gap, \emph{diagnostics} approaches aim to produce more informative verdicts.
In this paper, we address the problem of generating informative verdicts for violated Inter-procedural Control-Flow Temporal Logic (\icftl) specifications that express constraints over program variable values.
    We propose a diagnostic approach based on backward data-flow analysis to statically determine the relevant statements contributing to the specification violation.
    Using this analysis, we instrument the program to produce enriched execution traces.
    Using the enriched execution traces, we perform the runtime analysis and identify the statements whose execution led to the specification violation.
We implemented our approach in a prototype tool, \icftlDiagnostics, and evaluated it on \num{112} specifications across \num{10} software projects.
    Our tool achieves \SI{90}{\percent} precision in identifying relevant statements for \num{100} of the \num{112} specifications.
    It reduces the number of lines that have to be inspected for diagnosing a violation by at least \SI{90}{\percent}.
    In terms of computational cost, \icftlDiagnostics generates a diagnosis within \SI{7}{\minute}, and requires no more than \SI{25}{\mega\byte} of memory.
    The instrumentation required to support diagnostics
    incurs an execution time overhead of less than \SI{30}{\percent} and a memory overhead below \SI{20}{\percent}.

\end{abstract}

    \section{Introduction}
\label{sec:introduction}
\IEEEPARstart{M}{any} software systems operate in environments that are partially unknown or unpredictable at design time~\cite{sanchez2019survey}, making static verification techniques, like model checking, insufficient.
\emph{Runtime Verification (RV)}~\cite{bartocci2018introduction} has emerged as a lightweight formal method that monitors the behaviour of a system at runtime to determine whether it satisfies a given specification.
These specifications, often expressed in specification languages (such as MTL~\cite{mtl}, STL~\cite{stl}, CFTL~\cite{cftl}, iCFTL~\cite{dawes2021icftl}, SB-TemPsy-DSL~\cite{boufaied2020trace}), define properties over execution traces. 
The monitoring process yields a \emph{verdict}---typically Boolean or quantitative~\cite{deshmukh2017robust}---indicating whether the trace conforms or not to the given specification.
While indicating whether the system runtime behaviour satisfies the specification is useful, once a violation is detected, engineers require a deeper understanding of the underlying cause.
For example, consider a source-level specification such as: ``\emph{the value of variable \texttt{temp} in procedure \texttt{proc} should always be less than 10}.''
If a program execution violates this property, a conventional RV tool will report a Boolean verdict indicating failure.
However, in practice, engineers are interested in understanding what contributed to the violation---such as which operations or inputs influenced the value of \texttt{temp}.
This process of determining what led to the violation is known as \emph{diagnostics}, and it is tailored to the semantics of the specification language in use.
Diagnostics aim to explain \emph{why} a verdict was reached by identifying the violated subformulas and the relevant parts of the trace that contributed to the violation. 
For example, \citet{ferrere2015trace} proposed a method for extracting minimal subtraces that explain violations in MTL and STL specifications.
\citet{dawes2019explaining} presented a framework for identifying problematic segments of source-code that could explain a CFTL specification violation.
Similarly, \citet{boufaied2023trace} introduced a diagnostics approach for SB-TemPsy-DSL that annotates trace segments and state transitions to highlight the source of violations.

We address the diagnostics problem for \emph{iCFTL}, a specification language designed to express inter-procedural, source code-level properties of programs~\cite{dawes2021icftl}.
\icftl enables engineers to easily capture source code properties, making it suitable for software development workflows.
There are two types of \icftl specifications:
\begin{enumerate*}[label=(\roman*)]
    \item time-based: temporal constraints over program points, and
    \item state-based: constraints over program variables.
\end{enumerate*}
Our previous work~\cite{stratan2024diagnosing}, tackled the diagnostics problem for time-based properties by determining the unique event in the trace, called \emph{point of no return}, after which it is impossible to satisfy the specification.
Using this event, we identify the execution trace slice that contains the events that can explain the specification violation.
However, this approach does not generalise to \emph{state-based} properties since their violation is not tied to a specific moment in time but rather to how a variable’s value was computed.

In this work, we fill this gap by introducing a diagnostics approach tailored for \icftl \emph{state-based} properties, which focuses on data-flow analysis to trace how variables obtain their values across procedures and statements.
Such properties capture requirements like ``\emph{after the change of variable \texttt{temp} in procedure \texttt{proc}, variable \texttt{humidity} in procedure \texttt{proc} should have a value less than 20}'' or ``\emph{the values of variable \texttt{humidity} in procedure \texttt{proc} should always be less than the variable \texttt{temp} in procedure \texttt{proc}}.''
When encountering a violation of these specifications, we are interested in diagnosing how the variables \texttt{humidity} and \texttt{temp} obtained their values.
In this way, we enrich \icftl with full diagnostics capabilities for both time and state based properties.
To diagnose violations of state-based properties, we consider a single (inter-procedural) execution trace, denoted as an \iotatrace --- a sequence of events recorded during program execution, each tied to a specific program point within a procedure.
Diagnosing a single trace reflects practical constraints in real-world systems, where re-execution (to obtain multiple traces) may be costly or infeasible, especially in large-scale or production environments.
The goal of state-based diagnostics is to determine which events in the \iotatrace contributed to the violation of a specification.
We define an event as \emph{relevant} if it contributed to the computation of the value of any variable used in the specification.
As an example, consider the property ``\emph{the value of variable \texttt{temp} in procedure \texttt{proc} should always be less than 10},'' and two consecutive statements $x = 6$ and $temp = x + 5$.
Although the violation occurs at $temp = x + 5$, the assignment $x = 6$ is also relevant, as it influences the value of \texttt{temp}.
We propose an automated approach to identify such relevant events 
using backward data-flow analysis, traversing the procedure’s control flow graph, starting from the point of violation (i.e. the event that violates the specification).
During traversal, we collect all the nodes connected via backwards data-flow dependency to the point of violation, allowing us to trace how the procedure execution computed the variable's value that led to the violation.
Leveraging the inter-procedural capabilities of \icftl, we extend this analysis beyond individual procedures and capture data-flow dependencies across procedures boundaries.
Beyond identifying relevant events, we introduce a parameter called \emph{multiplicity}, which quantifies an event’s importance by counting the number of distinct data flow paths—such as those involving conditionals, loops, or control transfers (e.g., \texttt{break}, \texttt{continue})—linking it to the specification violation.
Intuitively, an event with higher multiplicity plays a more central role in the violation, as it influences the variable’s value through multiple execution paths.

We implemented our approach in a prototype tool, \icftlDiagnostics, by extending the original \icftl instrumentation approach to produce an enriched trace that includes events relevant to the specification violation.
This enriched trace forms the basis for computing the diagnosis, i.e., the set of relevant events and associated statements that explain the specification violation.
We empirically evaluated \icftlDiagnostics by diagnosing \num{112} specification violations for different procedures across \num{10} projects.
We address four distinct research questions (RQs), assessing \icftlDiagnostics effectiveness in identifying relevant statements (RQ1), its ability to reduce the complexity of understanding specification violations (RQ2), the time and memory required for diagnosis (RQ3), and the overhead introduced by its instrumentation approach (RQ4).
Our experimental results demonstrate that \icftlDiagnostics accurately identifies the events relevant to specification violations with at least \qty{90}{\percent} precision for 100 out of 112 specifications (RQ1).
In terms of complexity reduction (RQ2), \icftlDiagnostics significantly narrows down the portions of code that need to be inspected—by at least \SI{90}{\percent}—thereby facilitating the understanding of specification violations.
In terms of computational cost (RQ3), the diagnosis process completes within \SI{7}{\minute} and requires no more than \SI{25}{\mega\byte} of peak memory for the specifications we evaluated.
Furthermore, the additional instrumentation introduced by our tool incurs a reasonable overhead—less than \SI{30}{\percent} in execution time and under \SI{20}{\percent} in peak memory usage (RQ4).
To summarise, the main contributions of the paper are: 
\begin{itemize}[noitemsep, topsep=0pt, leftmargin=*]
    \item An approach based on backward data-flow analysis that identifies relevant events in an \iotatrace and computes a \emph{diagnosis} for the violated specification.
\item \icftlDiagnostics, a prototype implementation of our approach.
    \item An empirical evaluation assessing the effectiveness and efficiency of \icftlDiagnostics on \num{10} projects and \num{112} specifications.
\end{itemize}
The rest of the paper is structured as follows.
Section~\ref{sec:background} introduces the necessary \icftl background.
Section~\ref{sec:modif_scfg} discusses the modification we applied to how programs are represented in \icftl.
Section~\ref{sec:approach} describes our approach.
Section~\ref{sec:implementation} describes our prototype tool.
Section~\ref{sec:evaluation} reports on the experimental evaluation.
Section~\ref{sec:related-work} positions our contribution with respect to the literature.
Section~\ref{sec:concl-future-work} offers concluding remarks and a roadmap for future work.

An approach based on backward data-flow analysis that identifies relevant events in a trace and computes a diagnosis for the violated specification.
iCFTL-Diagnostics, a prototype implementation of our approach.
An empirical evaluation assessing the effectiveness and efficiency of iCFTL-Diagnostics on 10 projects and 112 specifications.     \section{Background: Inter-procedural Control-Flow Temporal Logic}
\label{sec:background}
In this section, we introduce a statically-computable representation of a program, followed by the notion of trace, i.e., the representation of a program's execution.
We then present the fragment of \icftl considered in this paper and its associated semantics.
All material introduced in this section is based on the original \icftl paper~\cite{dawes2021icftl} and our previous work~\cite{stratan2024diagnosing}.

\subsection{Statically-Computable Representation of a Program}
\subsubsection{Systems of Multiple Procedures}
We define a program as a \emph{system of multiple procedures} $\mathcal{S}$, which is a pair $\langle \mathcal{P},\mathit{prog}\rangle$ where $\mathcal{P}$ is a finite set of procedure names, and $\mathit{prog}$ is a map that sends each procedure name to its corresponding program.

\subsubsection{Symbolic Control Flow Graphs}
\label{subsec:scfg-intro}
In \icftl, for each procedure in $\mathcal{S}$, we compute a statically-computable representation, which we call \emph{symbolic control-flow graph} (SCFG).
Intuitively, this is a directed graph that uses vertexes to represent the \emph{symbolic} states reached by executing a statement in code.
Such \emph{symbolic states}, denoted by $\sigma$, indicate that a program variable's value may have changed, or a function may have been called, but encode no concrete values (since these are often only known at runtime).
For example, an assignment statement \texttt{x = a} would result in two symbolic states: one to represent the symbolic state of the program before \texttt{x} has been assigned a new value, and another to represent the symbolic state of the program in which \texttt{x} has just been assigned a value.
Formally, a symbolic state $\sigma$ is a pair containing the program point of a statement $\rho(stmt)$, and a map from program variables to one of the statuses $\{changed, unchanged, undefined, called\}$.
Since an SCFG is a directed graph, an edge from a symbolic state $\sigma$ to another symbolic state $\sigma'$ indicates that 
a statement is executed at runtime, transitioning the program to a new state.
Further, branching (caused by a conditional or loop) is represented by vertices having multiple successors.
Formally, for a procedure $p$, a symbolic control-flow graph, denoted by $\mathsf{SCFG}(p)$, is a triple $\langle V, E, v_s \rangle$ with $V$ being a set of \emph{symbolic states}, $E \subset V \times V$ being a set of edges, and $v_s$ being the initial symbolic state.
The initial symbolic state $v_s$ represents the entry point of the procedure—before any computation has occurred-and encodes the assignment of the procedure's parameters.
A path $\pi$ through an SCFG is defined as a sequence of edges $e_1,e_2,\ldots,e_n$, where each edge $e_i \in E$ (with $1\leq i\leq n$), and for every consecutive pair $e_i$ and $e_{i+1}$, we have $e_i = \langle \sigma, \sigma'\rangle$ and $e_{i+1} =\langle \sigma', \sigma''\rangle$; that is, the edges must be adjacent.

\subsection{Dynamic Runs}
We represent program executions as \emph{traces}, which we build by considering paths through SCFGs. 
Each trace consists of a sequence of SCFG vertices annotated with timestamps.
We use these annotated vertices as the basis of our notion of \emph{concrete states}, which are symbolic states augmented with information obtained at runtime (such as timestamps and program variables values).
Formally, a concrete state is a triple $\langle t, \sigma, m \rangle$ for $t$ a real-numbered timestamp, $\sigma$ a symbolic state, and $m$ a map from program variables to values observed at runtime.
We then represent executions of individual procedures as sequences of concrete states, which we call \emph{dynamic runs}.
Hence, a dynamic run $\mathcal{D}$ is a sequence $\langle t_1, \sigma_1, m_1 \rangle, \dots, $ $ \langle t_n, \sigma_n, m_n \rangle$ of concrete states such that there must be a path from each $\sigma_i$ to $\sigma_{i+1}$, for $1 \le i < n$, in the relevant SCFG.
Once concrete states have been collected into sequences, we refer to a pair of consecutive concrete states in a sequence as a \emph{transition}.
Intuitively, since concrete states correspond to symbolic states, we use pairs of concrete states (i.e., transitions) to model the computation that takes place at runtime to reach one concrete state from another. Concretely, we can use transitions to model operations like program variable value changes and function calls.

\subsubsection*{Inter-procedural Dynamic Runs}
We lift the notion of a dynamic run to model an execution of a system of multiple procedures $\mathcal{S}$ by labelling each dynamic run in a set with the name of the procedure to which it corresponds.
Formally, an \emph{inter-procedural dynamic run} $\mathcal{I}$ of $\mathcal{S}$ is a tuple $\langle \mathcal{P}, \{\mathcal{D}_1, \mathcal{D}_2, \dots, \mathcal{D}_n\}, \mathcal{L} \rangle$.
Here, $\mathcal{P}$ is the same $\mathcal{P}$ (set of procedure names) used to define $\mathcal{S}$,
$\{\mathcal{D}_1, \mathcal{D}_2, \dots, \mathcal{D}_n\}$ is a set of dynamic runs, and $\mathcal{L} : \{\mathcal{D}_1, \mathcal{D}_2, \dots, \mathcal{D}_n\} \to \mathcal{P}$ is a map that labels each dynamic run $\mathcal{D}_i$ with a procedure name from $\mathcal{P}$. For brevity, we will often refer to inter-procedural dynamic runs as $\iota-$traces.

To reason about parts of such $\iota-$traces, we introduce the notions of \emph{sub-trace} and \emph{sub-trace under a predicate} as defined in our previous work~\cite{stratan2024diagnosing}.
Intuitively, $\mathcal{I}'$ is a sub-trace of $\mathcal{I}$ if, when we take $\mathcal{I}$ and remove some concrete states, or dynamic runs, we obtain $\mathcal{I}'$.
More formally, we can define an injective function $\gamma$ that from a dynamic run in $\mathcal{I}'$ returns which dynamic run from $\mathcal{I}$ had concrete states removed to obtain it.
For example, given $\mathcal{I}=\langle \mathcal{P}, \{D_1,D_2\},\mathcal{L}\rangle$, with  $D_1=\langle t_1,\sigma_1,m_1\rangle,\langle t_2,\sigma_2,m_2\rangle,\langle t_3,\sigma_3,m_3\rangle$, and $\mathcal{I}'=\langle \mathcal{P}', \{D_1'\},\mathcal{L}'\rangle$, with $D_1'=\langle t_1,\sigma_1,m_1\rangle,\langle t_2,\sigma_2,m_2\rangle$, then $\mathcal{I}'$ is a sub-trace of $\mathcal{I}$ as $D_1'$ contains a subset of concrete states of $D_1$ and $\mathcal{I}'$ does not contain $D_2$.
Then, we can construct $\gamma$ as $\gamma(D_1')=D_1$.
As the sub-trace relation just states that we remove concrete states from dynamic runs but not which ones, we extend the sub-trace relation to require that any concrete state removed from the \iotatrace satisfies a given predicate.
We say that $\mathcal{I}'$ is \emph{a sub-trace under a predicate} $P$ of a trace $\mathcal{I}$, if
\begin{enumerate*}[label=(\roman*)]
    \item $\mathcal{I}'$ is a sub-trace of $\mathcal{I}$, and
    \item any concrete state $c$ found in the dynamic runs of $\mathcal{I}'$ must satisfy $P(c)=\mathit{true}$.
\end{enumerate*}
For example, a predicate $P$ could retain only concrete states that occurred within the first \SI{3}{\second} of system execution, i.e., $P(c) = \mathit{true}$ iff $t_c < 3$ with $c=\langle t_c, \sigma_c, m_c \rangle$.

\subsection{\icftl Specifications}
Here, we briefly introduce how \icftl specifications are composed and illustrate their semantics.
For a detailed definition of \icftl syntax and semantics we refer to~\cite{dawes2021icftl}.

Specifications in \icftl consist of \emph{quantifiers} and Boolean combinations of \emph{atomic constraints}. Quantifiers enable one to capture concrete states or transitions from $\iota-$traces, and bind them to variables. Atomic constraints enable one to place constraints over values extracted from concrete states and transitions. Specifically, atomic constraints are formed of \emph{expressions}, which enable one to select the concrete states or transitions to be used in the constraint. Further, \icftl specifications are in prenex normal form, and we assume that there are no free variables. For example, one can capture the property ``\emph{every time the program variable \texttt{y} is changed during the procedure \texttt{g}, the value of \texttt{y} during \texttt{g} should be less than 4 and the value of the next change of variable \texttt{x} after \texttt{y} should be less than 10}'' by writing
\begin{equation}
    \begin{split}
    &\forall q \in \texttt{changes}(y).\texttt{during}(g):\\
    &q(y) < 4 \ \land q.\texttt{next}(\texttt{changes}(x).\texttt{during}(g)) < 10
    \label{eq:spec-1}
    \end{split}
\end{equation}
This specification has one quantifier $ \forall q \in changes(y).during(g)$ and two atomic constraints, $q(y) < 4$ and $q.next(changes(x).during(g))<10$, connected by logical conjunction.
Each atomic constraint contains an expression, respectively, $q(y)$ (which is the notation for the value of variable $y$ at q) and $q.next(changes(x).during(g))$ (which is the updated value of variable $x$ after q).
Equation~\ref{eq:spec-1} serves as a template for the \emph{state-based} specifications that we consider in this work, which are specifications with one or more quantifiers that capture concrete states (variable value changes).
We now describe the key parts of this specification, as well as how the semantics would be used to compute a truth value, given an \iotatrace.

We begin with the quantifier, $\forall q \in changes(y).during(g)$, which looks for all concrete states representing
a change of the program variable \texttt{y}, during the execution of function \texttt{g}.
Formally, this involves
\begin{enumerate*}[label=(\roman*)]
    \item computing the SCFG of the procedure \texttt{g};
    \item inspecting the symbolic state $\sigma$ of each concrete state $c_i=\langle t, \sigma, m \rangle$ to check for a change of \texttt{y}.
\end{enumerate*}
For each concrete state $c_i$, the semantics constructs a \emph{binding} $\beta_i$, which maps $q$ (from the quantifier) to the concrete state $c_i$.
For each binding $\beta_i$, the semantics determines whether the atomic constraints $q(y) < 4$ and $q.next(changes(x).during(g)) < 10$ hold.
We first identify the unique concrete states to which the expressions $q(y)$ and $q.next(changes(x).during(g)) $ correspond.
For this, the $\mathsf{eval}(\mathcal{I}, \beta_i, \mathit{expr})$
function is used, which takes an \iotatrace $\mathcal{I}$, a binding
$\beta_i$, and an expression $\mathit{expr}$, and extracts the
concrete states relevant to that expression. Once the $\mathsf{eval}$ function returns the concrete state or transition that satisfies the expression, we can finally resolve the constraint.
First, we separately check each constraint for satisfaction, and then we compute the verdict for the whole specification.
The atomic constraint $q(y)<4$ is \texttt{true} if the value mapped to \texttt{y} in the concrete state returned by $eval$ is less than \num{4}.
Similarly, the atomic constraint $q.next(changes(x).during(g))<10$ is
\texttt{true} if the constraint (in the example, the value being less than \num{10}) is satisfied.
Then, the satisfaction of the specification for this binding is given by the conjunction of the values of the two atomic constraints.
Finally, the specification verdict is computed as a map, given by $[\mathcal{I},\varphi]_S$ that maps each binding to its truth value.

     \begin{figure}[t]
    \footnotesize
    \hspace{0.7cm}
    \noindent
    \begin{minipage}{0.35\linewidth}
\begin{lstlisting}[numbers=left,
                   frame=tlrb,
                   language=Python]
def k():
    a = 8
    for y in [0, 1]:
        b = a + 1
    endFor
    a = a + b
    m(a)
def m(a):
    c = 9
    g(c,a)
\end{lstlisting}
    \end{minipage}
    \hspace{1.3cm}
    \begin{minipage}{0.35\linewidth}
\begin{lstlisting}[numbers=left,
                   firstnumber=11,
                   showlines=true,
                   numberblanklines=false,
                   frame=tlrb,
                   language=Python,]
def g(b,y):
    l = y + 3
    if l < 8:
        k = l + 1
    else:
        k = b + l
    endIf
    y = k + 4
\end{lstlisting}
\end{minipage}
    \caption{Example of program with three procedures \texttt{k,m,g}. The execution of the program starts with procedure \texttt{k}, which calls procedure \texttt{m}, which eventually calls procedure \texttt{g}.}
    \label{fig:program-2}
\end{figure}

\section{Augmenting the SCFG with Dataflow Information}
\label{sec:modif_scfg}
In \S~\ref{subsec:scfg-intro}, we introduced the concept of SCFG, used in \icftl to statically represent procedures. 
While effective for capturing control flow, the original definition of SCFG lacks dataflow information about how variable values are computed, particularly in the context of state-based specifications.
This limitation prevents the detection of the program points that we want to include in the diagnosis of violated specifications.
For example, consider the program in Figure~\ref{fig:program-2}, and the specification ``\emph{for each change of variable \texttt{a} its value should always be less than \num{10}},'' which can be expressed in \icftl with the formula $\forall p \in changes(a).during(k): p(a) < 10$. 
For diagnosing the specification, we want to understand how the
variable \texttt{a} obtained its value at line~\num{6} (i.e., when the
value of \texttt{a} changes during the execution of \texttt{k}).
At line~\num{6}, the statement \texttt{a = a + b} shows that \texttt{a} depends on both \texttt{a} and \texttt{b}. 
However, in the SCFG, this dataflow dependency is not reflected, as the symbolic state for statement at line~\num{6} holds only information about the program point and the map from the variable \texttt{a} to one of the statuses: $\langle \rho_5(\mathit{stmt}), [a \mapsto changed]\rangle$. 

To address this limitation, we refine the SCFG by enriching symbolic states with additional dataflow information.
In the remainder of this paper, to distinguish between the two representations, we refer to the original SCFG as SCFG\textsuperscript{o}, and to the modified version—enriched with dataflow information—as simply SCFG.
In the modified SCFG, we incorporate dataflow information by tracing which variables are \emph{written}, and which variables are \emph{read} at each program point.
Furthermore, to enable the tracking of the dataflow across different procedures (i.e., inter-procedural analysis), we want to trace which functions are \emph{called} at each program point.\footnote{
    We chose the names \emph{written} and \emph{read}--rather than the naming \emph{defined} and \emph{used}---to be closer to the terminology of programming languages and reflect the concrete operations.
    This choice aligns with the \icftl goal of having a program representation---the SCFG---intuitive and directly mappable to the program’s syntax and execution behaviour.
}
To formalise this modification, we redefine the symbolic state to include information about which variables are read, written, and which functions are called.
Specifically, instead of recording only the status of the program variables with the map $m$, we use a triple $\langle \mathit{written}, \mathit{read}, \mathit{called}\rangle$, where: $\mathit{written}$ is the set of all the written variables at that program point $\rho(\mathit{stmt})$; $\mathit{read}$ is the set of variables whose values are read at that program point $\rho(\mathit{stmt})$; $\mathit{called}$ is the set of procedures that are called at that program point $\rho(\mathit{stmt})$.
\begin{figure}[t]
    \centering
    \includegraphics{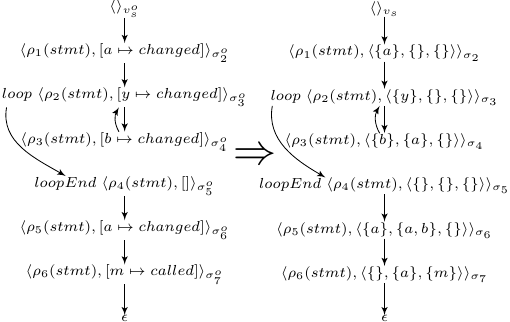}
    \caption{SCFG of procedure \texttt{k}: original (left) and new (right).}
    \label{fig:new-SCFG}
\end{figure}
For example, in the program of Figure \ref{fig:program-2}, the symbolic state corresponding to line \num{6} has the triple $\langle \mathit{written}=\{a\}, \mathit{read} = \{a,b\}, called=\{\}\rangle$.
We showcase the difference between SCFG\textsuperscript{o} and our modified SCFG using procedure \texttt{k} from Figure~\ref{fig:program-2}; both representations are shown in Figure~\ref{fig:new-SCFG}, with symbolic states indexed by the line number they correspond to.
To distinguish between the two representations, symbolic states in the SCFG\textsuperscript{o} are annotated with a superscript ``o'' (e.g., $\sigma^o_2$), while in the modified SCFG—and throughout the rest of the paper—we refer to the modified symbolic states simply as $\sigma$.
The SCFG\textsuperscript{o} captures only variable status changes (e.g., \texttt{a} is changed in $\sigma^o_6$), while on the right, the modified SCFG explicitly records which variables are written, read, and which functions are called at each program point. 
By extending symbolic states with this dataflow information, we can perform analysis to identify program points that influence a variable’s value.
For example, symbolic state $\sigma_4$ (where \texttt{b} is in $\mathit{written}$) can be now linked to the computation of \texttt{a} in $\sigma^o_6$. This was not possible in the SCFG\textsuperscript{o}, since $\sigma^o_6$ carried no information of \texttt{b} being used to compute \texttt{a}.

     \section{Approach}
\label{sec:approach}
\begin{figure*}[t]
    \centering
\includegraphics{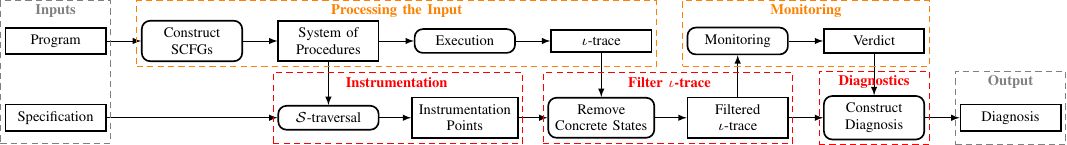}
    \caption{Overview of the diagnostic approach workflow. Rounded-corners and sharp-corners rectangles represent execution steps and artefacts, respectively.
    The red boxes highlight our contributions;
    the orange boxes indicate components from the original \icftl implementation.}
    \label{fig:approach-flow}
\end{figure*}

Our goal is to identify the concrete states in an \iotatrace that are useful to understanding how state-based atomic constraints were falsified in a specification and led to its overall violation.
Specifically, for each expression in the atomic constraint, we want to provide the set of
concrete states that are relevant to the constraint falsification, i.e.,  the concrete states that contributed to the computation of the expression value whose assignment led to the constraint violation.
Figure~\ref{fig:approach-flow} illustrates the overall workflow of our approach.
The approach takes as input a Python \texttt{Program}  and an \icftl \texttt{Specification} and produces the \texttt{Diagnosis} as output.
The key steps of the approach are:
\begin{enumerate}[label=(\roman*)]
    \item Processing the input, where we generate the SCFGs and
      \iotatrace;
    \item Instrumentation, where we detect the instrumentation points,
      i.e., the symbolic states potentially relevant to the
      specification violation;
    \item Filter \iotatrace, where we use the instrumentation points
      to filter the \iotatrace;
    \item Monitoring, where we obtain the verdict on the specification; and
    \item Diagnostics, where, in case of state-based specification
      violation, we retrieve the concrete states relevant to
      understanding the violated specification.
\end{enumerate}
Our contribution lies in the \texttt{Instrumentation}, \texttt{Filter \iotatrace} and \texttt{Diagnostics} steps (highlighted in red in Figure~\ref{fig:approach-flow}).
In the \texttt{Instrumentation} step (described in \S~\ref{subsec:instrumentation}), we traverse the SCFG and identify the symbolic states needed for explaining each of the expressions in the atomic constraints.
We refer to such symbolic states as ``instrumentation points''.
Using these instrumentation points, we filter the \iotatrace so that
it includes only concrete states relevant to each expression
(\texttt{Filter \iotatrace}, described in \S~\ref{subsec:filter-trace}).
We run the \icftl monitoring algorithm on the filtered \iotatrace to obtain the specification \texttt{Verdict} (\texttt{Monitoring}).
Finally, based on the verdict and \iotatrace, we construct the
diagnosis by first finding all the falsified atomic constraints, and
then constructing a diagnosis for each expression (\texttt{Construct
  Diagnosis}, described in \S~\ref{subsec:diagnostics}).

\begin{figure}[tb]
\begin{equation*}
\begin{split}
        \mathcal{I}_{ex} = &\langle \{k,m,g\}, \{D_k,D_m,D_g\}, \{k \mapsto D_k,m \mapsto D_m, \\
                                   &g \mapsto D_g\} \rangle
        \\
          \mathcal{D}_k = &\langle 0,[],[]\rangle, 
              \langle 0.2, \sigma_2,[a \mapsto 8]\rangle,
              \langle 0.4, \sigma_3, [y \mapsto 0]\rangle, \\
              & \langle 0.6, \sigma_4,[b \mapsto 9]\rangle,
              \langle 0.8, \sigma_3, [y \mapsto 1]\rangle, 
              \langle 1.0, \sigma_4,[b \mapsto 9]\rangle,\\
              & \langle 1.1, \sigma_5,[]\rangle,
              \langle 1.2, \sigma_6,[a \mapsto 17]\rangle,
              \langle 3.1,\sigma_7,[] \rangle
        \\
\mathcal{D}_m = &\langle 1.4,[],[]\rangle, 
          \langle 1.6,\sigma_9,[c \mapsto 9]\rangle, 
          \langle 2.9, \sigma_{10},[]\rangle
        \\
\mathcal{D}_g = & \langle 1.7,[],[]\rangle, 
          \langle 1.9, \sigma_{12},[l \mapsto 20] \rangle,
          \langle 2.0, \sigma_{13},[] \rangle, 
          \langle 2.2, \sigma_{15}, [] \rangle, \\
          & \langle 2.4, \sigma_{16},[k \mapsto 29] \rangle, 
          \langle 2.5, \sigma_{17},[] \rangle,
          \langle 2.7, \sigma_{18},[y \mapsto 23] \rangle
        \end{split}
    \end{equation*}
    \caption{Example of \iotatrace $\mathcal{I}_{ex}$ from the execution of the program in Figure~\ref{fig:program-2}.
    The trace is a triple of procedures, dynamic runs and a map from procedures to dynamic runs.
    The three dynamic runs $D_k, D_m$, and $D_g$ each represent the execution of one procedure.}
    \label{fig:trace}
\end{figure}

\subsection{Instrumentation}
\label{subsec:instrumentation}
The goal of the \emph{Instrumentation} step is to identify the set of
symbolic states in an \iotatrace such that the concrete states
connected to these symbolic states are sufficient for both trace checking and diagnostics.
An existing instrumentation approach for \icftl~\cite{dawes2021icftl}, referred to as \emph{vanilla instrumentation}, determines the minimal set of instrumentation points (i.e., symbolic states) sufficient to decide if an \iotatrace meets a given specification.
Building on the vanilla instrumentation, we extend the approach for diagnosing state-based specifications.
Starting from the symbolic states used for providing the verdict, we select the additional symbolic states needed to backtrack (diagnose) how the program execution led to the values of the variables relevant to the verdict.
Specifically, given a program and an expression, we aim to collect all
the symbolic states (both the one determined by the vanilla
instrumentation and the additional ones required for diagnostics) that contain information relevant, in terms of dataflow, to this expression.
We refer to this extended instrumentation that includes both vanilla and additional instrumentation points as \emph{diagnostics instrumentation}.

\paragraph{Running Example}
We showcase the difference in the instrumentation approaches with an example based on the program in Figure~\ref{fig:program-2} and the following \icftl specification:
\begin{equation}
    \forall q \in changes(y).during(g): q(y)<4 .
\label{eq:spec}
\end{equation}
This specification constraints the value of the variable \texttt{y} during the execution of procedure \texttt{g}.
The \icftl vanilla instrumentation scheme focuses on determining  symbolic states where \texttt{y} changes within SCFG(\texttt{g}), hence it will identify as instrumentation point only the symbolic state $\sigma_{18}$, associated with line \num{18}.
However, diagnostics instrumentation requires understanding how \texttt{y} obtained its value—specifically, which variables contributed to its computation.
In this case, we track \texttt{k}, which is directly used to compute the value of \texttt{y}.
Such a dataflow dependency is formally captured by the fact that, in the symbolic state $\sigma_{18}$, where \texttt{y} belongs to the \emph{written} set, \texttt{k} belongs to the \emph{read} set.
Reasoning recursively, we also want to understand how \texttt{k} obtained its values.
Thus, the diagnostics instrumentation includes also lines \num{16} and
\num{14}, as they assign a value to \texttt{k}\textemdash indeed, their associated symbolic states \texttt{k} belongs to the \emph{written} set.
Iterating further, we also track the values of \texttt{l} and \texttt{b}, to be able to diagnose how \texttt{k} obtained its value.
Accordingly, the instrumentation should include also lines \num{12}, \num{9}, \num{6}, \num{4} and \num{2}.
This iteration continues until all the symbolic states that are connected by dataflow to the assignment of \texttt{y} at line \num{18} are identified.
By statically instrumenting these symbolic states, one will be able to determine the chain of variable changes that led to the value of variable \texttt{y} that (potentially) violates the specification.
To rigorously identify these relevant states, we propose
Algorithm~\ref{alg:SCFG-traversal}, which we call \emph{system of
  procedures traversal}, shortened to  ``$\mathsf{\mathcal{S}\text{-}traversal}$.''

\begin{algorithm}[t]
  \caption{$\mathsf{\mathcal{S}\text{-}traversal}$}
    \label{alg:SCFG-traversal}
    \scriptsize
    \SetAlgoLined
    \textbf{Input:} $\sigma_n=\langle \rho_n(\mathit{stmt}),\langle \mathit{written_n}, \mathit{read_n}, \mathit{called}_n \rangle \rangle, \mathcal{U},  \mathcal{E}, \mathit{proc}$\\
    \KwResult{A new list $\mathcal{E}$ of symbolic states that explain $\sigma_n$}
    
    $\mathcal{E} : list \gets [\sigma_n]$\; \label{scfgTV:initE}
    $\mathcal{U} : set \gets \{\mathit{read_n}$\}; \label{scfgTV:initU}

    \SetKwProg{Def}{def}{:}{}
    \Def{$\mathsf{\mathcal{S}\text{-}traversal}$($\sigma_n$: symbolic state, $\mathcal{U}$: set of read variables, $\mathcal{E}$: list of additional instrumentation points, $\mathit{proc}$: procedure)}{
    $\mathit{incomingStar} = \mathit{getIncomingStar}(\sigma_n,\mathit{proc})$\; \label{scfgTV:incomingStar}
        \For{$\langle \rho_x(\mathit{stmt}),\langle \mathit{written}_x, \mathit{read}_x, \mathit{called}_x \rangle \rangle \in \mathit{incomingStar}$}
        {\label{scfgTV:iterateIncomStar}
        $\mathcal{U}'$ = copy of $\mathcal{U}$\; \label{scfgTV:copyU}
        $\mathscr{v}_s =  \langle \rho_s(\mathit{stmt}),\langle \mathit{written_s}, \mathit{read_s}, \mathit{called}_s\rangle\rangle$: starting symbolic state of $SCFG(\mathit{proc})$\;  \label{scfgTV:initSym}

\If{$\langle \rho_n(\mathit{stmt}),\langle \mathit{written_n}, \mathit{read_n}, \mathit{called}_n\rangle\rangle == \mathscr{v}_s$}{\label{scfgTV:ifInitSym}
            \For{$ w_s \in \mathit{written_s}$}{ \label{scfgTV:forWs}
            \If{$w_s \in \mathcal{U}'$}{ \label{scfgTV:ifWs}
$\mathit{callers} = getCallers(\mathit{proc},\mathcal{S})$\; \label{scfgTV:callers}
                $\mathit{parameterIndex} = \mathsf{getParameterIndex}(\mathscr{v}_s,w_s)$\; \label{scfgTV:parameterIndex}
                \For{$\mathit{caller} \in \mathit{callers}$}{ \label{scfgTV:forCallers}
                    $\mathit{name} = \mathit{caller}[0]$\; \label{scfgTV:callerName}
                    $\mathit{symbolicState} = \mathit{caller}[1]$\;  \label{scfgTV:callerSymState}
                    $\mathit{renamedParameter} = \mathsf{getRenamedParameter}(SCFG(\mathit{name}), $\\ \hspace*{2em}$
                    \mathit{symbolicState},
                    \mathit{parameterIndex})$\; \label{scfgTV:renamedParam}
                    $\mathsf{\mathcal{S}\text{-}traversal}(\mathit{symbolicState}, $\\ \hspace*{2em}$ \{\mathit{renamedParameter}\},\mathcal{E},\mathit{proc})$\;  \label{scfgTV:iterateOverParam}                            
                }
                $\mathcal{U}'= \mathcal{U}' \setminus \{w_s\}$\;  \label{scfgTV:removeWFromU}
            }}
        }

\If{$\mathcal{U}'\cap \mathit{written}_x==\emptyset$}{\label{scfgTV:ifParent}
            $\mathsf{\mathcal{S}\text{-}traversal}(\langle \rho_x(\mathit{stmt}),\langle \mathit{written}_x, \mathit{read}_x, \mathit{called}_x\rangle\rangle,\mathcal{U}',$\\ \hspace*{2em}$ 
            \mathcal{E},\mathit{proc})$\;  \label{scfgTV:iterateParents}
            \textbf{continue}\;
        }\label{scfgTV:endIfParent}

$\mathcal{E}.add (\langle \rho_x(\mathit{stmt}),\langle \mathit{written}_x, \mathit{read}_x, \mathit{called}_x\rangle\rangle)$\; \label{scfgTV:updateE}
        $\mathcal{U}'$ = $\mathcal{U}' \setminus \mathit{written}_x$\;  \label{scfgTV:updateU}
        \If{$\mathcal{U}' \neq \emptyset \lor \mathit{read}_x \neq \emptyset$}{ \label{scfgTV:ifUnotEmpty}
            $\mathcal{U}'$ = $\mathcal{U}' \cup \mathit{read}_x$\; \label{scfgTV:addLeftoverToU}
            $\mathsf{\mathcal{S}\text{-}traversal}(\langle \rho_x(\mathit{stmt}),\langle \mathit{written}_x, \mathit{read}_x, \mathit{called}_x\rangle\rangle,\mathcal{U}',$\\ \hspace*{2em}$
            \mathcal{E},\mathit{proc})$\;  \label{scfgTV:recursiveOnLeftover}
        }  \label{scfgTV:endIfUnotEmpty}
    }
    \Return $\mathcal{E}$\;  \label{scfgTV:return}
    }
\end{algorithm}

\subsubsection{$\mathsf{\mathcal{S}\text{-}traversal}$}
\label{subsubsec:scfg-traversal}
The $\mathsf{\mathcal{S}\text{-}traversal}$ algorithm, shown as pseudocode in Algorithm~\ref{alg:SCFG-traversal}, is applied to each symbolic state $\sigma_e$ that satisfies an expression of an atomic constraint in the specification\textemdash i.e., the symbolic states instrumented by the vanilla instrumentation.
It takes as input a symbolic state $\sigma_e=\langle\rho_e(\mathit{stmt}),\langle \mathit{written}_e, \mathit{read}_e, \mathit{called}_e\rangle\rangle$, a procedure $\mathit{\mathit{proc}}$, and a system of multiple procedures $\mathcal{S}$; it returns the list $\mathcal{E}$ of additional instrumentation points that are relevant to diagnosing the expression satisfied by the symbolic state $\sigma_e$.
This list of additional instrumentation points may have duplicates, since the same symbolic state may be relevant through different dataflow paths.
We track such duplicates to highlight symbolic states that have higher relevance for a given expression, as they can influence the verdict in multiple ways.

The algorithm performs a backward dataflow analysis to collect the list of {\em relevant} symbolic states that can affect the value of the variable in the expression.
This is done by recursively traversing the nodes in the SCFG starting from the symbolic state that satisfies the atomic constraint's expression.
At each step, we identify symbolic states that influence $\sigma_e$ either directly (if they assign a value to variable used directly in $\sigma_e$) or indirectly (if they assign a value to a variable further back in the dataflow dependency of $\sigma_e$).
More formally, given a procedure \texttt{proc}, its SCFG(\texttt{proc})=$\langle V, E, v_s\rangle$, and two symbolic states $\sigma_y=\langle \rho_y(stmt), \{\mathit{written}_y,\mathit{read}_y,\mathit{called}_y\} \rangle$ and $\sigma_x=\langle \rho_x(\mathit{stmt}), \{\mathit{written}_x,\mathit{read}_x,\mathit{called}_x\} \rangle$ in $V$,
\begin{definition}
    We say that $\sigma_y$ \emph{influences} $\sigma_x$, denoted by $\sigma_y\mathrel{\leadsto}\sigma_x$, if a variable from $\mathit{written}_y$ exists in $\mathit{read}_x$ and there exists a path $\pi$ in SCFG(\texttt{proc}) between $\sigma_y$ and $\sigma_x$:
    \begin{align*}
      \sigma_y\mathrel{\leadsto}\sigma_x \iff\
      & (\exists \mathit{var}: \mathit{var} \in \mathit{written}_y \land var \in \mathit{read}_x) \ \land \\
      & (\exists \pi=(e_1,...e_n): e_1,...e_n\in E : e_1=\langle\sigma_y,\ast\rangle  \\
      & \land e_n=\langle\ast,\sigma_x\rangle)
    \end{align*}
    where $\ast$ can be any symbolic state, and $n\geq1$.
    \label{def:influence}
\end{definition}
Then, the \emph{relevant} symbolic states in SCFG(\texttt{proc}) can be defined as follows:
\begin{definition}
    We say that $\sigma_y$ is \emph{relevant} to $\sigma_x$ if there exists a sequence of symbolic states $\sigma_1,...\sigma_n$ such that they sequentially influence each other, with $\sigma_1=\sigma_y$ and $\sigma_n=\sigma_x$, formally
    $
        \exists \sigma_1,...\sigma_n : \sigma_i\mathrel{\leadsto}\sigma_{i+1}\land \sigma_1=\sigma_y \land \sigma_n=\sigma_x
$
    where $1\leq i\leq n$.
    \label{def:relevant}
\end{definition}

To apply this definition and identify the additional instrumentation points, while traversing the SCFGs in the program, we track the $\mathit{read}$ variables of relevant symbolic states in the set $\mathcal{U}$, and the list of relevant symbolic states themselves in the list $\mathcal{E}$ (i.e., the instrumentation points).
We initialise list $\mathcal{E}$ with the symbolic state $[\sigma_e]$
that satisfies the expression (line~\ref{scfgTV:initE})
and $\mathcal{U}$ with its read variables (line~\ref{scfgTV:initU}).
At each symbolic state that we visit, we check if any of the variables in $\mathcal{U}$ is in its $\mathit{written}$ set \textemdash i.e., we check if it satisfies Definition~\ref{def:relevant}.
If so, the symbolic state is relevant to diagnosing the expression and should be instrumented.
Accordingly, we
\begin{enumerate*}[label=(\roman*)]
    \item add the symbolic state to  list $\mathcal{E}$
      (line~\ref{scfgTV:updateE}),
    \item remove from $\mathcal{U}$ the variables for which we found the definition of their value (line~\ref{scfgTV:updateU}), and
    \item add the $\mathit{read}$ variables of that symbolic state as
      they are relevant to the specification violation (lines~\ref{scfgTV:addLeftoverToU}).
\end{enumerate*}

To traverse backward the nodes of the program's SCFGs, we recursively explore them starting from the symbolic states that satisfy an expression of the atomic constraint.
For each visited node we inspect its parents in the SCFG, which we refer to as its \emph{incoming star}.
When computing the incoming star of a node (function \emph{$\mathit{getIncomingStar}$} at line~\ref{scfgTV:incomingStar}), we have three main cases depending on the control flow of the program.
\begin{itemize}[noitemsep, topsep=0pt, leftmargin=*]
    \item If there is no control flow, the incoming star will contain only one incoming statement.
    \emph{Example}: In the program in Figure~\ref{fig:program-2}, the incoming star of the symbolic state $\sigma_{13}$ (line 13) is $\sigma_{12}$ (line 12).
    \item In case the incoming symbolic state is the end of an \texttt{if} statement, we want to separately iterate over both paths of the conditional as we do not know which path will be executed at runtime.
    Thus, we will have two symbolic states in the incoming star, one for each branch of the conditional.
    \emph{Example}: Considering again the program in Figure~\ref{fig:program-2}, the incoming star of the symbolic state $\sigma_{18}$ (line 18) contains both $\sigma_{14}$ and $\sigma_{16}$ (lines 14 and 16).
    \item In case the incoming star of the symbolic state contains a \texttt{loop} statement, we want to avoid infinite iterations over the loop body.
    Accordingly, we traverse the loop body only once and then, if needed, continue to the symbolic states before the loop.
    Thus, in the incoming star of the symbolic state right after the loop we include only the symbolic state corresponding to the last statement in the loop body (and not the symbolic state before the loop).
    Conversely, in the incoming star of the first symbolic state in the loop body we include only the symbolic state before the loop (and not the symbolic state at the end of the loop body).
    \emph{Example}: In program in Figure~\ref{fig:program-2}, the incoming star of $\sigma_6$ (line 6) is only $\sigma_4$ (line 4), and the incoming star of $\sigma_3$ (line 3) is only $\sigma_2$ (line 2).
\end{itemize}
We then iterate over the symbolic states in the incoming star (line~\ref{scfgTV:iterateIncomStar}).
Since each symbolic state represents a different dataflow path, we use different copies of $\mathcal{U}$ to avoid interference (line~\ref{scfgTV:copyU}).
If the symbolic state is not relevant—formally, when the intersection $\mathcal{U}' \cap \mathit{written}_x$ is empty—we perform a recursive call to continue exploring the program (lines~\ref{scfgTV:ifParent}–\ref{scfgTV:endIfParent}). 
Otherwise, we have found a relevant symbolic state: we update $\mathcal{E}$ by adding the current symbolic state and refine $\mathcal{U}$ by removing the variables that have now been explained (lines~\ref{scfgTV:updateE}–\ref{scfgTV:updateU}).
Finally, we are left with handling the inter-procedural aspect of the program.
This appears when a symbolic state of the incoming star is the starting symbolic state ($v_s$) of a procedure's SCFG (checked in lines~\ref{scfgTV:initSym}--\ref{scfgTV:ifInitSym}).
In such a symbolic state, the $\mathit{written}_s$ set contains the procedure parameters.
If any of the variables in $\mathit{written}_s$ is in $\mathcal{U}$ (checked in lines~\ref{scfgTV:forWs}--\ref{scfgTV:ifWs}), the dataflow continues outside of the procedure that we are currently exploring.
Then, we continue exploring the program where the procedure was called.
This requires:
\begin{enumerate*}[label=(\roman*)]
    \item finding where the current procedure $\mathit{proc}$ was called (line~\ref{scfgTV:callers}), and
    \item updating the name of the variable in $\mathcal{U}$ to the one used in each of the calling procedures (lines~\ref{scfgTV:parameterIndex}--\ref{scfgTV:removeWFromU}).
\end{enumerate*}

\begin{algorithm}[t]
  \caption{Function to find callers of a procedure in a program}
    \label{alg:getCallers}
    \scriptsize
    \SetAlgoLined
    \textbf{Input:} $\mathit{proc}, \mathcal{S}$\\
    \KwResult{A set of symbolic states and corresponding procedures that called $\mathit{proc}$}
      
    \SetKwProg{Def}{def}{:}{}
    \Def{getCallers($\mathit{proc}$: procedure, $\mathcal{S}$: system of multiple procedures)}{
    $\mathit{callers} : set \gets \{\}$\;

    \For{$p \in \mathcal{S}$}{
        \For{$\langle \rho_x(\mathit{stmt}),\langle \mathit{written}_x, \mathit{read}_x, \mathit{called}_x \rangle \rangle \in SCFG(p) $ }{
            \If{$\mathit{proc} \in \mathit{called}_x$}{
                $\mathit{name} = p$\;
                $\mathit{symbolicState} = \langle \rho_x(\mathit{stmt}),\langle \mathit{written}_x, \mathit{read}_x, \mathit{called}_x \rangle \rangle$\;
                $\mathit{callers} = \mathit{callers} \cup \langle \mathit{name},\mathit{symbolicState}\rangle$\;
            }
        }
    }
    \Return $\mathit{callers}$\;
    }
\end{algorithm}

To find the calls to $\mathit{proc}$ in the system of procedures $\mathcal{S}$, we use Algorithm~\ref{alg:getCallers}.
This algorithm takes as input a procedure $\mathit{\mathit{proc}}$ and system
of procedures $\mathcal{S}$; it returns the set $\mathit{\mathit{callers}}$ of procedures that contain a call to $\mathit{\mathit{proc}}$ together with the symbolic state where the call happens.
The algorithm iterates over the procedures in $\mathcal{S}$ (line~\num{4}); for each symbolic state in the SCFG of that procedure (line~\num{5}) it checks if procedure $\mathit{\mathit{proc}}$ is in the set of called procedures (line~\num{6}).
When such symbolic states are found, we add the procedure name and
symbolic state to $\mathit{\mathit{callers}}$ (line \num{9}).

Back to Algorithm~\ref{alg:SCFG-traversal}, after computing the
$\mathit{callers}$ at line~\ref{scfgTV:callers}, it proceeds to find the name used for the variable in $\mathcal{U}$ in them.
First, we identify which parameter of the function it corresponds to.
We do so through by calling the $\mathsf{getParameterIndex}$ function
(line~\ref{scfgTV:parameterIndex}), which identifies which of the positional arguments is the one corresponding to the variable in $\mathcal{U}$.
Then, we iterate over the $\mathit{callers}$ (line~\ref{scfgTV:forCallers}).
For each call to $\mathit{proc}$, we find the name of the variable in
$\mathcal{U}$ based on the parameter's position
(lines~\ref{scfgTV:forCallers}--\ref{scfgTV:renamedParam}), and
recursively call $\mathsf{\mathcal{S}\text{-}traversal}$ over the symbolic state of
the function call ($\mathit{symbolicState}$), the caller procedure
($\mathit{name}$) and with $\mathcal{U}$ re-initialised to the new
variable, while keeping the same $\mathcal{E}$ (line~\ref{scfgTV:iterateOverParam}).
Finally, we can remove $w_s$ from the set of used variables $\mathcal{U}$ (line~\ref{scfgTV:removeWFromU}).
The recursion of $\mathsf{\mathcal{S}\text{-}traversal}$ stops when either $\mathcal{U}$ is empty (i.e., we are able to fully diagnose the specification), or the incoming star is empty (i.e., there are no more parts of the program to explore).
Finally, the list of instrumentation points $\mathcal{E}$ is returned (line~\ref{scfgTV:return}).
\begin{figure}[t]
    \centering
    \begin{subfigure}{0.98\columnwidth}
        \centering
        \includegraphics{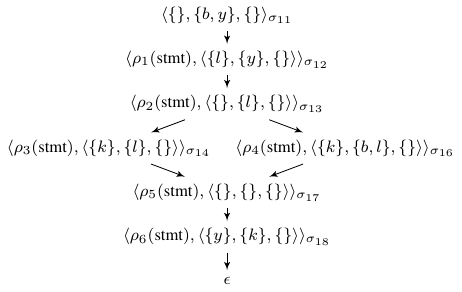}
        \caption{SCFG of procedure \texttt{g} from the program in Figure~\ref{fig:program-2}. Each node represents a symbolic state, denoted by $\sigma_i$ with an index $i$  indicating the corresponding line number. The entry state of the procedure is $\sigma_{10}$ and the exit state is $\epsilon$.}
        \label{fig:scfg-only-if-else}
    \end{subfigure}
    \begin{subfigure}{0.98\columnwidth}
        \centering
        \vspace{2mm} \includegraphics{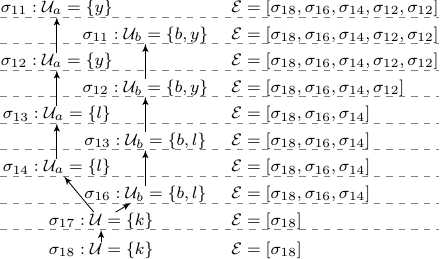}
        \caption{Changes of $\mathcal{U}$ and $\mathcal{E}$ when applying Algorithm~\ref{alg:SCFG-traversal} to SCFG(g). 
        We showcase the branching of the paths and how  $\mathcal{U}$ and $\mathcal{E}$ are computed in that case.
        In each node, the left side shows the symbolic state being processed and the updated $\mathcal{U}$; for each path we have a different copy of $\mathcal{U}$.
        On the right side we show the updated $\mathcal{E}$ after processing that symbolic state.}
        \label{fig:scfg-traversal-if-else}
    \end{subfigure}
    \caption{Example of applying Algorithm~\ref{alg:SCFG-traversal} to procedure $g$.}
    \label{fig:scfg-if-else}
\end{figure}

\paragraph{Application to the running example}
We illustrate the working of $\mathsf{\mathcal{S}\text{-}traversal}$ using the example program in Figure~\ref{fig:program-2} alongside the specification in Equation~\ref{eq:spec}.
$\mathsf{\mathcal{S}\text{-}traversal}$ is called over the symbolic state $\sigma_{18}$ of procedure \texttt{g}.
We first illustrate the intra-procedural part of $\mathsf{\mathcal{S}\text{-}traversal}$ execution over the SCFG of \texttt{g}.
Then, we showcase the inter-procedural part by describing the identification and traversal of the callers of \texttt{g}.

To illustrate the intra-procedural aspect, we use Figure~\ref{fig:scfg-if-else}, where Subfigure~\ref{fig:scfg-only-if-else} shows SCFG(\texttt{g}) and Subfigure~\ref{fig:scfg-traversal-if-else} shows the recursive calls to the $\mathsf{\mathcal{S}\text{-}traversal}$ algorithm.
In the latter, each node represents a call to Algorithm~\ref{alg:SCFG-traversal}.
The graph is to be read bottom-up (as shown by the arrows) to reflect the backward traversal of the SCFG.
For each call (i.e., for each node of the graph), we show the updated
set $\mathcal{U}$ of used variables. Differently, for the updated list $\mathcal{E}$, since it is shared across all algorithm calls, we report it on the right as it gets updated by each algorithm call.

The traversal begins at node $\sigma_{18}$ (as mentioned above, this is the symbolic state that satisfies the expression in Equation~\ref{eq:spec}), where $\mathcal{U} = \{k\}$ is initialised with the variables read at $\sigma_{18}$, and $\mathcal{E}$ is initialised as $[\sigma_{18}]$.
The incoming star of $\sigma_{18}$ contains the symbolic state $\sigma_{17}$, which represents the statement marking the end of the conditional (i.e., \texttt{endIf}).
The analysis continues with a recursive call on $\sigma_{17}$.
The incoming star of $\sigma_{17}$, being after a conditional branching, contains two symbolic states: $\sigma_{14}$ and $\sigma_{16}$.
Thus, the algorithm is (recursively) called twice to independently explore the two paths, as represented by the branching of the graph in Figure~\ref{fig:scfg-traversal-if-else}.
Importantly, the used-variable set is copied for each of the two calls: $\mathcal{U}_a$ for the \texttt{if} branch ($\sigma_{14}$) and $\mathcal{U}_b$ for the \texttt{else} branch.
This is needed as, after the two calls, in the \texttt{if} branch, only $l$ is used to compute the value of $k$, resulting in an updated $\mathcal{U}_a=\{l\}$, while in the \texttt{else} branch both $b$ and $l$ are used, resulting in an updated $\mathcal{U}_b=\{b,l\}$.
The two calls then recurse both over $\sigma_{13}$, where there is no assignment and thus both continue to $\sigma_{12}$.
Here, both calls add $\sigma_{12}$ to $\mathcal{E}$ (note the
duplicate in the list), as it is needed for the explanation of $l$, found both in $\mathcal{U}_a$ and $\mathcal{U}_b$.
Ultimately, both calls reach $\sigma_{11}$, one with $\mathcal{U}_a=\{y\}$ and the other with  $\mathcal{U}_b=\{b,y\}$.

At this point, the $\mathsf{\mathcal{S}\text{-}traversal}$ algorithm transitions to the inter-procedural analysis and identifies the call site of \texttt{g} within procedure \texttt{m}.
The algorithm performs inter-procedural analysis twice---once for each branch--- by separately exploring the definitions of variables in $\mathcal{U}_a$ and $\mathcal{U}_b$.
Before performing the recursive call on procedure \texttt{m} and the symbolic state where \texttt{g} is called, the sets are updated as follows: 
$\mathcal{U}_a$ is updated to $\{a\}$, as the variable \texttt{y} is named \texttt{a} in the caller; analogously, $\mathcal{U}_b$ is updated to $\{c,a\}$.
When traversing procedure \texttt{m}, in the symbolic state $\sigma_9$ there is an assignment to the variable $c$, found in $\mathcal{U}_b$.
Thus, this symbolic state is added to the list of instrumentation
points, which is now $\mathcal{E} = [\sigma_{18},\sigma_{16},
\sigma_{14}, \sigma_{12}, \sigma_{12}, \sigma_9]$, and $c$ is removed
from the set $\mathcal{U}_a$ of $\mathit{read}$ variables. Finally, to resolve the definition of variable \texttt{a}, the two
calls recurse on $\sigma_9$ in function \texttt{k}, where \texttt{m}
is called. Ultimately the $\mathsf{\mathcal{S}\text{-}traversal}$ algorithm concludes by returning the list $\mathcal{E}= [\sigma_{18},\sigma_{16},\sigma_{14},\sigma_{12}, \sigma_{12},\sigma_9,\sigma_6,\sigma_6,\sigma_4,\sigma_4,\sigma_2,\sigma_2]$.
In the final list, we highlight the duplicates of $\sigma_6$, $\sigma_4$, and $\sigma_2$, generated by the two separate recursive calls caused by the branching at line~\num{13}.

\subsubsection{Multiplicity}
\label{subsubsec:multiplicity}
As mentioned above, the list of symbolic states $\mathcal{E}$ obtained using Algorithm~\ref{alg:SCFG-traversal} might contain duplicates, in the case that a symbolic state is relevant to a specification through multiple dataflow paths.
Accordingly, we want to compute the multiplicity of each symbolic state as a measure of its relevance for the specification.
To compute the multiplicity, we represent $\mathcal{E}$ as a multiset, i.e. a pair $(\mathcal{M}_{\mathcal{E}},\mu_{\mathcal{E}})$, where $\mathcal{M}_{\mathcal{E}}=\{\sigma \mid \sigma \in \mathcal{E}\}$ is a set and $\mu_{\mathcal{E}}:\mathcal{M}_\mathcal{E} \implies \mathbb{N}$ is a map representing the multiplicity of elements in $\mathcal{M}_\mathcal{E}$.
For each element $\sigma \in \mathcal{M}_\mathcal{E}$, its multiplicity is computed as:
$
    \mu_\mathcal{E}(\sigma) = \sum_{x \in \mathcal{E}} \delta(\sigma, x)
$
where $\delta(y, z)$ is the Kronecker delta function which values 1 when $y=z$ and 0 when $y\neq z$.
For example, given the list ${\mathcal{E}}= [\sigma_{18},\sigma_{16},\sigma_{14},\sigma_{12}, \sigma_{12},\sigma_9,\sigma_6,\sigma_6,\sigma_4,\sigma_4,\sigma_2,\sigma_2]$, the set $\mathcal{M}_{\mathcal{E}} $ is $\{\sigma_{18},\sigma_{16},\sigma_{14}, \sigma_{12},\sigma_9,\sigma_6,\sigma_4,\sigma_2\}$, and the multiplicity function $\mu$ is defined as $
\mu(\sigma_{18}) = 1,\,
\mu(\sigma_{16}) = 1,\,
\mu(\sigma_{14}) = 1,\,
\mu(\sigma_{12}) = 2,\,
\mu(\sigma_9) = 1,\,
\mu(\sigma_6) = 2,\,
\mu(\sigma_4) = 2,\,
\mu(\sigma_2) = 2
$.
Thus, the multiset is $(\{\sigma_{18},\sigma_{16},\sigma_{14}, \sigma_{12},\sigma_9,\sigma_6,\sigma_4,\sigma_2\}, \mu)$ with the multiplicity function $\mu$ providing the count of each element in $\mathcal{E}$.

\subsection{Filter \iotatrace}
\label{subsec:filter-trace}
In this step, we use the set of instrumentation points
$\mathcal{M}_{\mathcal{E}}$ to filter an \iotatrace $\mathcal{I}$, and
obtain a filtered \iotatrace that contains only the concrete states
necessary for trace checking and diagnostics.
We use the notion of sub-traces under a predicate to define an \fslice of an \iotatrace $\mathcal{I}$.
\begin{definition}
    We say that $\mathcal{I}'$ is an \fslice of $\mathcal{I}$ if, $\mathcal{I}'$ is a sub-trace of $\mathcal{I}$ under the predicate
\begin{equation}
    P(\langle t,\sigma,m\rangle) = 
    \begin{cases} 
    \mathit{true} & \text{if } \sigma \in \mathcal{M}_{\mathcal{E}}\\
    \mathit{false} & \text{if } \sigma \notin \mathcal{M}_{\mathcal{E}} 
    \end{cases},
\label{eq:predicate}
\end{equation}
where $\mathcal{M}_\mathcal{E}$ is a set of instrumentation points.
\label{def:f-slice}
\end{definition}
We now introduce how we take a \iotatrace $\mathcal{I}$ and compute a \fslice.
Given an \iotatrace $\mathcal{I} = \langle \mathcal{P}, \{\mathcal{D}_1, \dots, \mathcal{D}_n\}, \mathcal{L} \rangle$ and the predicate $P$ over concrete states (from Equation~\ref{eq:predicate}), we compute the \fslice $\mathcal{I}' = \langle \mathcal{P}', \{\mathcal{D}'_1, \dots, \mathcal{D}'_k\}, \mathcal{L}' \rangle$ as follows:
\begin{enumerate}[label=(\roman*)]
    \item We compute the filtered set of dynamic runs by traversing each original dynamic run $ \mathcal{D}_i \in \{\mathcal{D}_1, \dots, \mathcal{D}_n\}$, and retaining only the concrete states $c$ for which $P(c) = \mathit{true}$.
    Each non-empty original dynamic run becomes a new dynamic run $\mathcal{D}'_j \in \{\mathcal{D}'_1, \dots, \mathcal{D}'_k\}$.
    Further, we record the mapping $\gamma(\mathcal{D}'_j) = \mathcal{D}_i$ to link each filtered dynamic run from the \fslice to its original in the \iotatrace.
    \item We compute the set of procedures $ \mathcal{P}' $ in the \fslice as $\mathcal{P}' = \{ p : \exists \mathcal{D}' \in \text{dom}(\gamma) : \mathcal{L}(\gamma(\mathcal{D}'))  = p \} $; that is, for each filtered dynamic run $ \mathcal{D}' \in  \{\mathcal{D}'_1, \dots, \mathcal{D}'_k\}$, we identify its dynamic run from $\mathcal{I}$ via $ \gamma(\mathcal{D}')$ and deduce the corresponding procedure by using the original mapping $\mathcal{L}$ from $\mathcal{I}$.
    \item We define the map $\mathcal{L}' $ such that $\mathcal{L}'(\mathcal{D}') = p$ if and only if the original dynamic run $\gamma(\mathcal{D}')$ was mapped to procedure $p$ in $\mathcal{L}$.
\end{enumerate}
Consequently, the \fslice retains only the concrete states that satisfy predicate $P$ in Equation~\ref{eq:predicate}, and builds the procedure set and map according to the remaining dynamic runs.
We compute the \fslice for each expression in the atomic constraints of the specification, using the set of instrumentation points $\mathcal{M}_{\mathcal{E}}$ of each expression (computed during the instrumentation step, described in \S~\ref{subsec:instrumentation}).

\paragraph{Application to the running example}
Continuing our example from \S~\ref{subsubsec:multiplicity}, we use the set $\mathcal{M}_{\mathcal{E}}= \{\sigma_{18}, \sigma_{16}, \sigma_{14}, \sigma_{12}, \sigma_9, \sigma_6, \sigma_4, \sigma_2\}$ to filter the \iotatrace $\mathcal{I}_{ex}$ in Figure~\ref{fig:trace}.
In the \iotatrace, we identify all the concrete states across the dynamic runs that are also in $\mathcal{M}_{\mathcal{E}}$, and discard the rest.
The resulting \fslice $\mathcal{I}'_{ex}$ consist of the filtered dynamic runs
$\mathcal{D}_k' = \langle 0.2, \sigma_2, [a \mapsto 8] \rangle, \langle 0.6, \sigma_4, [b \mapsto 9] \rangle, \langle 1.0, \sigma_4, [b \mapsto 9] \rangle, \langle 1.2, \sigma_6, [a \mapsto 17] \rangle$,
$\mathcal{D}_m' = \langle 1.6, \sigma_9, [c \mapsto 9] \rangle$, 
$\mathcal{D}_g' = \langle 1.9, \sigma_{12},[l \mapsto 20] \rangle,\langle 2.4, \sigma_{16}, [k \mapsto 29] \rangle, \langle 2.7, \sigma_{18}, [y \mapsto 23] \rangle$. 
Notably, there is no concrete state corresponding to $\sigma_{14}$, as it belongs to a conditional branch that was not taken during execution.

\begin{algorithm}[t]
\caption{computeMap}
\label{alg:diagnostics-map}
\scriptsize
\SetAlgoLined
\textbf{Input:} An \icftl specification $\varphi$ and an \iotatrace $\mathcal{I}$\\
\KwResult{A map from binding/expression pairs to f-slices of $\mathcal{I}$}

$\mathsf{diagnosticMap} \gets []$\;
$\mathsf{bindings} \gets \{\beta : [\mathcal{I}, \varphi]_S(\beta) = \mathsf{false}\}$\;
\For{binding $\beta \in \mathsf{bindings}$} {
    $\mathsf{falsifyingAtomicConstraints} \gets \mathsf{getFalsifyingAtomicConstraints}(\mathcal{I}, \beta, \varphi)$\;
    \For{constraint $\alpha$ in $\mathsf{falsifyingAtomicConstraints}$} {
        \For{expression $e$ in constraint $\alpha$} {
            $f\text{-}slice \ \langle \mathcal{P}',  \{\mathcal{D}'_1,  \dots,
            \mathcal{D}'_n\}, \mathcal{L} \rangle \gets getF\text{-}Slice(\mathcal{I}, \beta, e)$\;
            $\mathit{annotatedFSlice} \gets []$\;
            \For{dynamic run $\mathcal{D}'$ in $ \{\mathcal{D}'_1,  \dots,
            \mathcal{D}'_n\}$}{

            \For{concrete state $\langle t,\sigma,m\rangle$ in $\mathcal{D}'$}{
                $\mathit{annotatedFSlice} \gets \mathit{annotatedFSlice} \cup \langle\langle t,\sigma,m\rangle, \mu(\sigma)\rangle$\;
            }}
                $\mathsf{diagnosticMap} \gets \mathsf{updateMap}(\mathsf{diagnosticMap}, \langle \beta, e \rangle, \mathit{annotatedFSlice})$\;
        }
   
    }
}
\Return $\mathsf{diagnosticMap}$\;
\end{algorithm}

\subsection{Diagnostics}
\label{subsec:diagnostics} 
Using the concept of \fslice $\mathcal{I}'$ from the previous section (Definition~\ref{def:f-slice}), in this step we construct the diagnosis for a falsified specification.
More precisely, we construct a diagnosis:
\begin{enumerate}[label=\roman*]
    \item for each falsified binding of the specification (i.e., each time the specification is assessed over the trace), and
    \item for each expression that led to the negative value of the binding.
\end{enumerate}
Accordingly, a diagnosis is a map from pairs of falsified binding and expression, to the \fslice containing the concrete states relevant to that expression.

To compute this diagnostics map from pairs of falsified bindings and expressions to their corresponding \fslice, we apply Algorithm~\ref{alg:diagnostics-map} to the \icftl specification and the \iotatrace.
This algorithm iterates over the falsified bindings (line 4).
For each binding, it retrieves the set of atomic constraints that
falsify it (line \num{5}).
Specifically, function $\mathsf{getFalsifyingAtomicConstraints}$ takes an \iotatrace $\mathcal{I}$, a binding $\beta$ and a specification $\varphi$, and returns the set of atomic constraints whose truth value resulted in $[\mathcal{I}, \varphi]_S(\beta)$ being $\mathsf{false}$.
Then, for each falsifying constraint (line 6) and for each expression in it (line 7), we retrieve the \fslice containing only the concrete states relevant for the diagnosis (line \num{8}).
More precisely, function $\mathsf{getF\text{-}Slice}$ takes the \iotatrace $\mathcal{I}$, the binding $\beta$ and the expression $e$ in the atomic constraint $\alpha$ and retrieves the corresponding \fslice (see \S~\ref{subsec:filter-trace}).
To include the multiplicity of each of the concrete states in the \fslice, we expand the \fslice to an $\mathit{annotatedFSlice}$ holding such information (line \num{12}).
Finally, $\mathsf{updateMap}$ (line \num{13}) takes the map
$\mathit{diagnosticMap}$, along with the key $\langle\beta, e \rangle$
and the 
value $\mathit{annotatedFSlice}$ and updates the map to have $\mathsf{diagnosticMap}(\langle\beta, e \rangle) = \mathit{annotatedFSlice}$.

\paragraph{Application to the running example}
Let us consider again the program in Figure~\ref{fig:program-2} and the specification in Equation~\ref{eq:spec}.
In the program's execution trace in Figure~\ref{fig:trace}, the quantifier in the specification has a single binding $\beta$ from $q$ to $\langle 2.7, \sigma_{18},[y \mapsto 23] \rangle$ (the only change of variable \texttt{y} during the execution of $g$).
We retrieve the \fslice $\mathcal{I}'_{ex}$ from \S~\ref{subsec:filter-trace} example and annotate it with multiplicity information.
We compute the annotated \fslice ${\mathcal{I}'}_{ex}^A$, which consists of the following dynamic runs:
\begin{align*}
\mathcal{D}_k' = &\langle \langle 0.2, \sigma_2, [a \mapsto 8] \rangle, 2\rangle, 
            \langle\langle 0.6, \sigma_4, [b \mapsto 9] \rangle, 2\rangle, \\
            & \langle\langle 1.0, \sigma_4, [b \mapsto 9] \rangle, 2\rangle, 
            \langle\langle 1.2, \sigma_6, [a \mapsto 17] \rangle, 2\rangle,\\[4pt]
\mathcal{D}_m' = &\langle\langle 1.6, \sigma_9, [c \mapsto 9] \rangle, 1\rangle,\\[4pt]
\mathcal{D}_g' = &\langle\langle 1.9, \sigma_{12},[l \mapsto 20] \rangle,2\rangle,
            \langle\langle 2.4, \sigma_{16}, [k \mapsto 29] \rangle, 1\rangle, \\
            & \langle \langle 2.7, \sigma_{18}, [y \mapsto 23] \rangle, 1\rangle.
\end{align*}
Finally, we construct the diagnostic map $(\beta,q(y))\mapsto {\mathcal{I}'}_{ex}^A$.

     \section{Implementation}
\label{sec:implementation}
We have implemented our diagnostics approach in a prototype tool named \icftlDiagnostics.
Our tool builds upon a prior extension for diagnosing time-based specifications~\cite{stratan2024diagnosing}, which in turn is based on the existing framework for instrumentation and trace checking with respect to \icftl~\cite{dawes2021icftl}.
To support state-based specifications diagnostics, we further extended this framework by modifying both the instrumentation component and the \iotatrace filtering algorithm. 
Below, we describe the state-based diagnostics enhancements made to the extended \icftl infrastructure and list the limitations of our tool implementation.

\paragraph{Diagnostics instrumentation}
We further modified the existing
instrumentation, by extending the diagnostics instrumentation to
include program points relevant to state-based specifications (as
described in \S~\ref{subsec:instrumentation}).

\paragraph{Diagnostics}
We further extended the diagnostics capabilities of \icftlDiagnostics to diagnose state-based specifications, by extending the verdict to the map discussed in \S~\ref{subsec:diagnostics}.
We integrated the diagnostics component after the \texttt{Monitoring} process. 
Given the \texttt{Verdict} and \texttt{Filtered \iotatrace}, we apply the diagnostics approach and compute a diagnosis for each expression in the violated specification.
Each concrete state included in the diagnosis is annotated with its multiplicity, indicating through how many distinct execution paths it contributed to the specification violation.

\paragraph{Limitations}
The implementation of our approach inherited the known limitations of the original implementation of \icftl~\cite{dawes2021icftl} that it builds upon: 
\begin{itemize}[noitemsep, topsep=0pt, leftmargin=*]
    \item Global variables: The implementation of the tool does not instrument global variables, which may lead to missed concrete states when the analysis involves concrete states defined or modified outside the programs' procedures.
    \item Class Instantiations and Class-level Variables: The implementation of the tool does not instrument program points within class-level constructs, such as object instantiations, and class member variables.
    \item Array/List Element Definitions: Arrays and list are treated as a ``whole,'' and definitions or updates to individual elements within arrays or lists are not tracked.
    \item Complex Function Parameters: The implementation of the tool does not fully resolve nested or transformed parameters within function calls. For example, in the function call \texttt{move(a, int(b))}, the implementation will fail to recognize that \texttt{int(b)} references the variable \texttt{b}, potentially omitting concrete states from the \iotatrace relevant to the computation of its value.
\end{itemize}
We will discuss the impact of these limitations in the next section, when analysing the
experimental results.

     \section{Evaluation}
\label{sec:evaluation}
In this section, we report on the experimental evaluation of our
\icftlDiagnostics tool.
We focus on the effectiveness of the tool in identifying the causes of
state-based specification violations, as well as on its efficiency.
Regarding the \emph{effectiveness}, we assess the tool ability to identify
statements that are relevant to a given specification (i.e., the
statements that can affect the value of the variables used in a
specification), as well as the reduction---in terms of statements to
inspect---of the effort for identifying the cause
of specification violations when the tool output is used.
As for the \emph{efficiency}, we assess the time taken and the memory
used by our diagnostics approach, as well as the overhead introduced
by the corresponding instrumentation.
We summarise these objectives with the following research questions:
\begin{enumerate}[label=\bfseries RQ\arabic*:, leftmargin=3em]
    \item How effective is \icftlDiagnostics in identifying the statements relevant to a specification violation?
    \item To what extent does \icftlDiagnostics reduce the complexity of understanding specification violations?
    \item How much time and memory does the \icftlDiagnostics approach require?
    \item What is the overhead in terms of time and memory consumption introduced by the \icftlDiagnostics instrumentation? \end{enumerate}

\begin{table}[t]
    \centering
    \tiny
\begin{tabular}{p{1.2cm}p{1.1cm}p{0.55cm}p{0.245cm}p{0.245cm}lp{0.245cm}p{0.245cm}l}
\hline
{\bf Project Name} & {\bf Domain} & {\bf \# Specs} & \multicolumn{3}{c}{\bf SLOC} & \multicolumn{3}{c}{\bf \# Functions}\\
\hline
{}                 & {}            & {}       & {\bf Min} & {\bf Max}  & {\bf Avg}    & {\bf Min} & {\bf Max} & {\bf Avg}   \\ 
PythonRobotics     & Robotics      & \num{92} & \num{53}  & \num{454}  & \num{193.85} & \num{4}   & \num{33}  & \num{12.66} \\
flask\_api         & RESTful API   & \num{2}  & \num{52}  & \num{97}   & \num{74.50}  & \num{4}   & \num{7}   & \num{5.50}  \\
seaborn            & Data visual.  & \num{7}  & \num{410} & \num{2497} & \num{1944}   & \num{26}  & \num{74}  & \num{58.71} \\
pyfilesystem2      & Files         & \num{4}  & \num{70}  & \num{677}  & \num{410.75} & \num{6}   & \num{61}  & \num{36.50} \\
pypdf              & Format proc.  & \num{2}  & \num{317} & \num{481}  & \num{399}    & \num{20}  & \num{32}  & \num{26}    \\
pdoc               & Documentation & \num{1}  & \num{341} & \num{341}  & \num{341}    & \num{21}  & \num{21}  & \num{21}    \\
pyjwt              & Authentic.    & \num{1}  & \num{109} & \num{109}  & \num{109}    & \num{12}  & \num{12}  & \num{12}    \\
arrow              & Date \& time  & \num{1}  & \num{868} & \num{868}  & \num{868}    & \num{71}  & \num{71}  & \num{71}    \\
pudb               & Debugging     & \num{1}  & \num{443} & \num{443}  & \num{443}    & \num{18}  & \num{18}  & \num{18}    \\
elasticsearch\_dsl & Search        & \num{1}  & \num{45}  & \num{45}   & \num{45}     & \num{6}   & \num{6}   & \num{6}     \\
\hline
\end{tabular}

     \caption{List of Selected Project from the DyByBench benchmark.}
    \label{tab:projects}
\end{table}

\subsection{Datasets and Settings}
\subsubsection{Evaluation Subjects}
We evaluated  \icftlDiagnostics using subjects from different domains.
As the tool supports Python programs, we looked for a benchmark written in this language and including test cases (so that we
could assess our approach on realistic executions of the programs).
Under these constraints, we selected the DyByBench benchmark~\cite{bouzenia2024dypybench}, which encompasses 50 Python open-source projects from various application domains, with multiple test cases for each project. 
Notably, the benchmark includes projects for which the test cases provide high code coverage.
Such a high coverage is important to assess our tool, as it allows us
to assess our tool ability to trace relevant statements through different parts of the program.

The assessment of \icftlDiagnostics requires to define state-based specifications, i.e., specifications that constrain the values assigned to numerical variables, like integers and floats.
Thus, among the projects in the benchmark, we selected those
containing programs over which we could write state-based
specifications (leaving us with \num{26} projects).
However, due to the limitation of the underlying \icftl
implementation, we could not write specification over elements of arrays and lists.
Consequently, we excluded \num{16} projects where numerical variables appeared
only in arrays and lists.
After filtering out these projects, we were left with a total of 10
projects, listed in Table~\ref{tab:projects} together with their
application domain and the number of specifications (\# Specs) written
for the programs in each of the projects.
Additionally, for the programs within these projects over which the specifications were written, we provide the minimum, maximum, and average values of Source Lines of Code (SLOC) and the number of functions (\# Functions);
we computed the SLOC using the \texttt{Radon} tool~\cite{radonAPI}
to count non-comment, non-blank lines, and used the Python \texttt{ast} module~\cite{pythonAST}
to traverse the abstract syntax tree and count function definitions.

\subsubsection{Specifications}
To operate our tool on diverse portions of the programs, for each of the 10 selected projects, we wrote each specification for a different procedure in them;
this ensures different instrumentation points for the different specifications.
When writing the specifications, we targeted variables in procedures executed by the test cases.
Furthermore, to write specifications relevant to the intended purpose of the procedure under test, when possible, we based them on the oracles available with the test cases.
For example, if a test case contained an assertion statement over a variable, such as \texttt{assert(len(x)==3)}, we included this assertion in the specification.
Finally, to ensure that the trace diagnostics be triggered, we wrote
specifications that would be violated (i.e., would provide a negative verdict) by the test execution.
The specifications contain either a single atom or a Boolean combination of atoms.
For example, a single atom specification looks like 
$    \forall q \in changes(x).during(g):q(x)<10$, while a specification with multiple atoms looks like $\forall q \in changes(x).during(g): q(x)< 0 \land q.next(changes(y).during(g)>10)$.
In total, we wrote 112 specifications for the different functions in the 10 selected projects from the benchmark.

\subsubsection{Settings}
All experiments were conducted on the Aion High-Performance Computing
system at the University of Luxembourg.\footnote{Experiments were conducted on a node equipped with an AMD EPYC ROME 7H12 processors (\num{64} cores, \SI{2.6}{\giga\hertz}, \SI{280}{\watt}) and \SI{128}{\giga\byte} of RAM. Further system specifications are available at: \url{https://hpc-docs.uni.lu/systems/aion/}.}

We did not consider any baseline because, to the best of our knowledge, no existing tool provides diagnostics for \emph{state-based} specifications in the context of \icftl or similar specification languages. 

\subsection{RQ1: Effectiveness of \icftlDiagnostics}
To assess the effectiveness of our tool in identifying statements
relevant to a specification violation, we compared the instrumentation points
generated by \icftlDiagnostics
for the evaluation subjects (and their respective specifications) with
a manually defined ground truth (GT).
We created this GT by manually inspecting the code and computing an explanation set for each specification.
Since the GT is computed statically, we do not have access to the execution traces.
Therefore, we evaluate effectiveness with respect to the statically determined instrumentation points produced by the \icftlDiagnostics approach.

\subsubsection{Ground Truth Definition}
The GT is the set of program points that can affect the outcome verdict of a given specification (i.e., the set of relevant program points).
We created this GT by manually inspecting the code and computing the instrumentation points for each specification.

While all the identified program points can affect the specification outcome, their impact on the specification outcome can vary.
In particular, program points that are closer---in terms of backward dataflow dependency---to where the specification is violated are expected to have a more direct influence on the outcome.
Intuitively, missing a program point that directly assigns a value to a variable used in the specification is more detrimental than missing one that assigns a value to a variable not directly used in the specification.
More rigorously, while all the program points in the GT are connected by a dataflow dependency with at least one of the program points that can violate the specification, each of them can be located in a different procedure of the program.
Accordingly, missing a program point close (in terms of dependency) to where the specification is violated is more detrimental than missing one that is further away in the dependency chain.

To capture this distinction, we assign a value, which we call \emph{proximity}, to each program point in the GT.
We define the proximity value based on the function that the program point belongs to: it measures the distance from the function where the specification was violated to the function the program point is in.
Each program point receives a proximity value equal to \num{0} if it is in the same function where the specification was violated; \num{1} if it is in a function that calls the function at proximity \num{0};  \num{2} if it is in a function that calls a function at proximity \num{1}; and so on.
While we define proximity for any inter-procedural distance, we also
define proximity for relevant program points outside of any procedure
in the program (e.g., global and class variables) for which we set
the proximity value to  $\infty$.
In the special case that a program point is connected by backward
dependency with multiple points where the specification might be
violated and thus it can have multiple proximity values, we select the
lowest one.
We exemplify the proximity values using the program points in the diagnosis of the specification in Equation~\ref{eq:spec} for the program in Figure~\ref{fig:program-2}.
As mentioned above, the quantifier in the specification is satisfied by the symbolic state $\sigma_{18}$.
Among the symbolic states identified in our diagnosis (see \S~\ref{subsec:diagnostics}), proximity \num{0} includes $\sigma_{16}$, $\sigma_{14}$, and $\sigma_{12}$, as they are within the same procedure as $\sigma_{18}$.
Proximity \num{1} includes $\sigma_9$, as it is located in procedure \texttt{m}, which directly calls procedure \texttt{g}.
The remaining program points within procedure \texttt{k} are at proximity \num{2}, since \texttt{k} calls \texttt{m}, which in turn calls \texttt{g}.
Finally, in this example, there are no program points with assigned proximity of $\infty$, i.e., global or class variables.
For our evaluation subjects, the GT (including the proximity values) was first created by the first author, then validated independently by the second author, and finally discussed by the two until convergence.

\subsubsection{Methodology}
For each specification, we measured the effectiveness of our tool by comparing the GT with the instrumentation points generated by the tool itself.
We performed this comparison for different proximities, grouping program points cumulatively based on their proximity level:
for a given proximity $n$, we included program points having a proximity value ranging from \num{0} to $n$.
When comparing at proximity $\infty$, we compared the instrumentation points with the GT set including all program points.
If, for a given specification and proximity, the two subsets coincided, i.e., they contained the same program points, then we concluded that our tool identified {\em all} the relevant statements.
The more the two subsets differed, the least effective we deemed our tool.

As such, we compare the GT set with the \icftlDiagnostics instrumentation points set $\mathcal{M}_\mathcal{E}$ for a given proximity. 
To compare the sets, we considered as a True Positive (TP) a program point in  $\mathcal{M}_\mathcal{E}$ that is part of the GT, a False Positive (FP) a program point in  $\mathcal{M}_\mathcal{E}$ that is not part of the GT, and a False Negative (FN) a program point in the GT that is not present in $\mathcal{M}_\mathcal{E}$.
We then computed {\em precision} $\precision=\frac{\mathit{TP}}{\mathit{TP} + \mathit{FP}}$ and
{\em recall} $\recall=\frac{\mathit{TP}}{\mathit{TP} + \mathit{FN}}$.
Precision measures the percentage of the identified program points that are actually relevant to the specification violation.
Recall instead measures the percentage of relevant statements that our tool was able to identify.
Using this definition, to answer our RQ, we executed our tool on the 112 specifications, obtained the corresponding $\mathcal{M}_\mathcal{E}$ set and computed precision and recall at different proximities.

\begin{figure}
    \centering
    \scriptsize
    \includegraphics{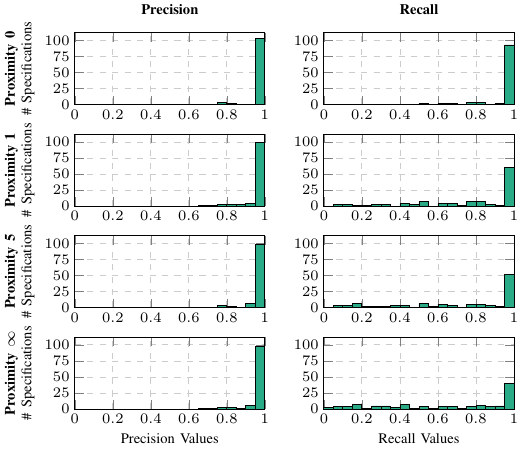}
    \caption{Histograms of precision (left) and recall (right) values obtained when comparing the GT and the $\mathcal{M}_\mathcal{E}$ sets. 
In each row, we report the results for increasing proximity values.
    High precision and recall values indicate overlap between the two sets, and therefore good performance of our tool.
    }
    \label{fig:rq1}
\end{figure}

\begin{figure}
    \centering
    \scriptsize
    \includegraphics{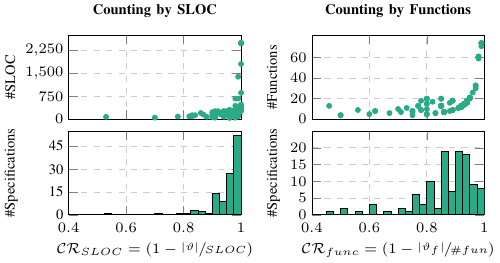}
    \caption{The plots show the reduction of complexity quantified in SLOC ($\mathcal{CR}_{SLOC}$, left) and functions ($\mathcal{CR}_{func}$, right).
    }
    \label{fig:rq2}
\end{figure}

\subsubsection{Results}
We report the results of our experiments in Figure~\ref{fig:rq1} as histograms of the precision and recall values over the \num{112} specifications.
In the figure, the histograms in the left-hand column show the precision values, and the ones in the right-hand column show the recall values.
The different rows instead show the results for proximity values \num{0},\num{1},\num{5} and $\infty$.
In terms of precision, our results show high values across all
proximity values, with at least \num{103} out of \num{112} specifications
having a precision above \num{90}\%.
This gives us confidence that the vast majority of program points detected by \icftlDiagnostics are indeed relevant.

In terms of recall, we observe that for proximity \num{0}, \num{95} out
of \num{112} specifications have a recall higher than \num{90}\%,
indicating that \icftlDiagnostics detects the vast majority of
relevant program points within the function where the specification
violation happens. When we consider higher proximity values, the recall values decrease for some of the specifications.
We have \num{61} specifications with recall higher than \num{90}\% for proximity \num{1}, \num{53} specifications for proximity \num{5}, and \num{45} specifications for proximity $\infty$.
Upon inspection, we observed that the missing statements leading to a
reduction of the recall value are due to the known limitations of
\icftlDiagnostics.
Notably, \icftlDiagnostics does not instrument global variables, class instantiations and class-level variables, array/list elements, and complex function parameters.
For instance, global variables assignments can be part of the dataflow path leading to a specification violation and therefore are included in the GT.
However, since these statements are not part of any procedure, they are not detected by \icftlDiagnostics, resulting in missing statements, as well as any other statements relevant to them.
This limitation appears at higher proximity levels, contributing to the observed decrease in recall for proximity $\infty$.
The complete list of such limitations is reported in \S~\ref{sec:implementation}.
\begin{Sbox}
\begin{minipage}{0.95\linewidth}
    {\bf RQ1 answer.} \icftlDiagnostics identifies the program
    points relevant to the specification violation with high precision 
    across different proximity values.
    Furthermore, our experiments also show high recall at proximities
    \num{0} and \num{1}, indicating that the tool detects the vast majority of relevant program points in the function where the specification is violated and in the functions that call it.
    While the recall decreases for higher proximities ($>$\num{1}), due to the known limitations of the tool implementation, this is less critical than missing program points closer to the specification violation.
\end{minipage}
\end{Sbox}
\shadowbox{\TheSbox}

\subsection{RQ2: Reducing Complexity in Understanding Violations}
To answer RQ2, we measured the complexity of understanding a specification violation in terms of the code that needs to be inspected after observing a violation.
The idea is that, compared to using no tool, \icftlDiagnostics leads to the inspection of a \emph{smaller} portion of the program for understanding the specification violation. 

\subsubsection{Methodology}
We quantified this reduction of complexity at two levels of granularity: one fine-grained, measured in terms of lines of code, and one more coarse-grained, measured in terms of functions.
At the fine-grained level, without any tool support, all of the lines of the program need to be inspected as they can potentially affect the specification verdict.
As such, the ones that actually affect the verdict have to be manually identified.
In contrast, when using \icftlDiagnostics, the diagnosis automatically identifies and provides the lines that directly contributed to the violation of the specification.
Then, the understanding of the violation cause can start directly from those lines.
Therefore, we measured the reduction of complexity as the reduction in lines that need to be inspected thanks to the use of \icftlDiagnostics.
For the coarse-grained assessment of the reduction of complexity, we compared instead the number of functions containing relevant program points identified by \icftlDiagnostics with the total number of functions in the program.
Like for the fine-grained assessment, the idea is that, without tool support, all the program functions need to be inspected as they potentially affect the specification verdict.
Differently, with \icftlDiagnostics, it is sufficient to inspect only those functions that contain program points included in the set of instrumentation points.

\begin{figure}[t]
\includegraphics{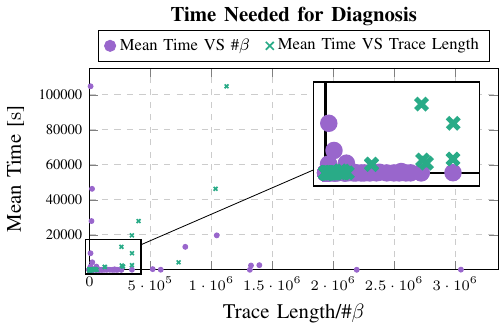}
    \caption{This figure shows the computational time required by our diagnosis step for our evaluation subjects.
    For each dot of the scatter plot, the vertical coordinate is the time required for performing the diagnosis, the horizontal coordinate is the number of bindings for the purple dots, and trace length for the green crosses.}
    \label{fig:rq3-time}

\includegraphics{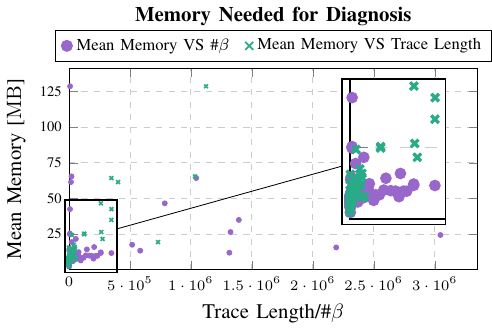}
    \caption{This figure shows the peak memory required by our diagnosis step for our evaluation subjects.
    For each dot of the scatter plot, the vertical coordinate is the peak memory required for performing the diagnosis, the horizontal coordinate is the number of bindings for the purple dots, and trace length for the green crosses.}
    \label{fig:rq3-memory}
\end{figure}

In practice, for each of our 112 specifications, we computed the fine-grained reduction of complexity as $\mathcal{CR}_{\mathit{SLOC}}=\left(1 - \sfrac{|\vartheta|}{\mathit{SLOC}}\right)$, where $|\vartheta|$ is the cardinality of the diagnosis set $\vartheta$, and $\mathit{SLOC}$ is the total number of Source Lines of Code (SLOC) in the program.
We computed the coarse-grained reduction of complexity as $\mathcal{CR}_{\mathit{func}}=\left(1 - \sfrac{|\vartheta_f|}{\mathit{\#fun}}\right)$, where $|\vartheta_f|$ is the cardinality of the set of functions containing at least one program point in the diagnosis set, and $\mathit{\#fun}$ is the total number of functions in the program.
For example, in a program of 100 SLOC and 5 functions, if \icftlDiagnostics gives a diagnosis of 25 statements over 3 functions, we obtain a reduction in complexity of $\mathcal{CR}_{\mathit{SLOC}}=\left(1 - \frac{25}{100}\right)=0.75$ and $\mathcal{CR}_{\mathit{func}}=\left(1 - \frac{3}{5}\right)=0.4$.
This means that the tool reduces the amount of code one has to analyse by 75\% and the number of functions by 40\%.

\subsubsection{Results}
We report the $\mathcal{CR}_{\mathit{SLOC}}$ and $\mathcal{CR}_{\mathit{func}}$ results of our experiments in Figure~\ref{fig:rq2}.
For both metrics, we include a scatter plot illustrating the reduction of complexity (on the x-axis) in relation to the total number of SLOC and of functions (on the y-axis), respectively.
In the histograms, we show the distribution of the reduction of complexity values  obtained for the 112 specifications.
We group specifications that have similar values of complexity reduction.
In terms of SLOC, the histogram shows that for 102 out of the \num{112} specifications, our tool desirably reduces the complexity of understanding the specification violation by more than \qty{90}{\percent}.
This indicates that our tool reduces the need to analyse the entire program from the total number of SLOC to just about \qty{10}{\percent} of it.
Similarly, in terms of functions, we observe that 100 specifications show more than \qty{80}{\percent} complexity reduction.
This indicates that \icftlDiagnostics reduces the need to analyse the entire program from the total number of functions to just about \qty{20}{\percent} of it.
In the two scatter plots, we observe that programs with a higher number of SLOC or functions tend to have a higher reduction in complexity.
This shows, intuitively, that \icftlDiagnostics is more beneficial
when applied to larger programs.
\begin{Sbox}
\begin{minipage}{0.95\linewidth}
    {\bf RQ2 answer.} Our experiments show that \icftlDiagnostics can reduce the number of SLOC and functions that need to be inspected to understand a specification violation by more than \qty{90}{\percent} and by \qty{80}{\percent}, respectively.
\end{minipage}
\end{Sbox}
\shadowbox{\TheSbox}

\subsection{RQ3: Efficiency of \icftlDiagnostics}
To answer this RQ,  we measured the time and memory taken by the
\icftlDiagnostics tool to analyse the traces generated by our evaluation subjects for each specification.

\subsubsection{Methodology}
To measure the time cost, we used the \texttt{time} module from the Python standard library to obtain the wall clock time.
We then computed the difference between the clock value before and after running the diagnosis approach algorithm to obtain its time cost.
To measure the memory cost, we used the \texttt{tracemalloc} module, which allows one to trace memory allocations.
We enabled and disabled the tracing at the start and end of the diagnostics execution, and measured the {\em peak size} of allocated memory.
By measuring the peak memory allocation, we obtain an upper bound of the memory required by the \icftlDiagnostics tool.
To account for the time and memory consumption variability across different executions, we executed the diagnosis five times and calculated the average time and memory.
Furthermore, to avoid interferences between the time and memory measurements, we used separate runs for the two quantities.

For each program and specification pair, the diagnostics is triggered by the occurrence of a falsified binding.
Furthermore, the diagnostics cost is influenced by the number of events to be processed, which depends on the length of the trace.
Since the number of bindings and trace length quantities can vary over different program and specification pair, we also studied the time and memory cost in relation to them.
Our \num{112} specifications show a wide variety of trace length values and number of bindings.
The trace length ranged from \num{1} to \num{1124250}, with an average of \num{54930.71}.
The number of bindings ranged from \num{3} to \num{3046663}, with an average of \num{140088.16}.
This diversity allows us to assess our diagnostics approach on diverse application contexts.

\subsubsection{Results}
We report our experiments results for the time and memory cost in Figures~\ref{fig:rq3-time} and~\ref{fig:rq3-memory}, respectively.
In each figure, we show the computational cost  of each specification
in relation to the trace length (TL) with green crosses, and in relation
to the number of bindings (\#$\beta$) with purple dots.

\begin{figure}
    \centering
    \includegraphics{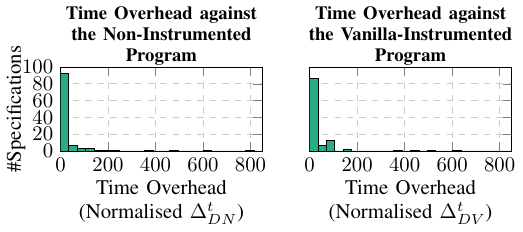}
    \caption{The program execution time overhead generated by \icftlDiagnostics in our evaluation subjects.
    }
    \label{fig:rq4-time}

\includegraphics{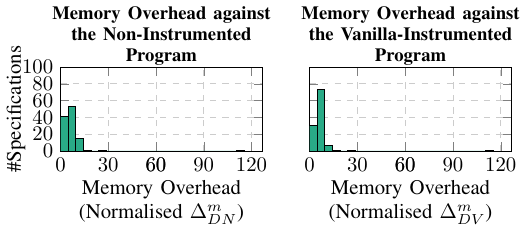}
    \caption{The program execution peak-memory overhead generated by \icftlDiagnostics in our evaluation subjects.
    }
    \label{fig:rq4-memory}
\end{figure}

In Figure~\ref{fig:rq3-time}, we observe that most of the dots/crosses are grouped in the bottom part of the plot, desirably showing short diagnosis times.
Specifically, the majority of the specifications require less than \SI{400}{\s}, or approximately \SI{7}{\min}.
The green crosses, which indicate the relation with the TL, show that the time taken to yield a diagnosis increases with the TL.
Differently, the purple dots, which indicate the relation with the number of bindings \#$\beta$, do not show any particular relation with the time cost.
Such qualitative observations are quantitatively confirmed by the Pearson coefficient, which is a normalised covariance of two variables.
A coefficient value close to \num{1} corresponds to a linear (positive) relation between the two variables, a value close to \num{-1} corresponds to a linear (negative) relation, and a value close to \num{0} corresponds to no correlation.
Between time and TL we observe a value of $\mathit{corr}^t_{\mathit{TL}}=0.85$, which indicates a strong linear correlation.
Between time and \#$\beta$ we observe a value close to
$\mathit{corr}^t_{\#\beta}=0.01$, which indicates low correlation.

In Figure~\ref{fig:rq3-memory}, we observe that the majority of the specifications desirably require less than \SI{25}{\mega\byte} of peak memory.
Differently, in terms of time cost, the memory cost increases as both TL
and \#$\beta$ increase.
This correlation is confirmed by the Pearson correlation coefficients.
Between memory usage and TL, we observe a coefficient value of $\mathit{corr}^m_{\mathit{TL}}=0.88$, indicating a strong positive correlation.
Instead, between memory usage and \#$\beta$, the coefficient is
lower, with $\mathit{corr}^m_{\#\beta}=0.24$.
This low value indicates that an increase in \#$\beta$ has limited impact on the memory usage.
\begin{Sbox}
\begin{minipage}{0.95\linewidth}
    {\bf RQ3 answer.} The computational cost of \icftlDiagnostics
    is low, requiring less than \SI{7}{\min} of time for the diagnosis and
    a peak memory usage of less than \SI{25}{\mega\byte} for processing our specifications.
    This indicates that \icftlDiagnostics is usable in practice, also for
    specifications that yield trace length (TL) and number of bindings (\#$\beta$) up to \num{500000}.
    Furthermore, our results show that while higher trace length values increase
    both time and memory costs, a higher number of bindings leads only to
    a higher peak-memory cost, without increasing the computational time.
\end{minipage}
\end{Sbox}
\shadowbox{\TheSbox}

\subsection{RQ4: \icftlDiagnostics Instrumentation Overhead}
We answer this RQ by measuring the execution time and memory overhead that our new instrumentation approach introduces on program execution.
As our work builds on top of the iCFTL~\cite{dawes2021icftl} framework, we study the program-execution overhead generated by \icftlDiagnostics both in absolute terms, comparing with the non-instrumented program, and relative terms, comparing with the basic \icftl instrumentation without any diagnostics functionality (hereafter referred to as  vanilla instrumentation).

\subsubsection{Methodology}
Similar to RQ3, to measure the peak memory cost, we used the \texttt{tracemalloc} module.
To measure the execution time of each test case, we used the Unix \texttt{time} command.
We computed the two overheads as the time and memory delta between the
diagnostics instrumentation and both the vanilla instrumentation and
the non-instrumented program.
To ensure comparability across different specifications, we normalised these differences, specifically:
\begin{itemize}
    \item The overhead relative to the vanilla instrumentation was divided by the vanilla time and memory costs.
    \item The overhead relative to the non-instrumented program was divided by the corresponding non-instrumented time and memory costs.
\end{itemize}
For each specification, we run the associated test case triggering the execution of the program under the different instrumentation schemes.
Like RQ3, we run separate experiments for time and memory.
To account for the variability of the individual executions, we executed the test cases 5 times each and compute the mean of both quantities.

\subsubsection{Results}
We report the overhead introduced by our instrumentation on program execution in Figure~\ref{fig:rq4-time} for time and in Figure~\ref{fig:rq4-memory} for memory.
In both figures, the histograms on the left show the diagnostics
instrumentation (D) overhead with respect to the non-instrumented program execution (N), which we denote as $\Delta^t_{\mathit{DN}}$ for time and $\Delta^m_{\mathit{DN}}$ for memory.
The histograms on the right show the diagnostics instrumentation (D) overhead with respect the vanilla instrumented program execution (V), which we denote as $\Delta^t_{\mathit{DV}}$ and $\Delta^m_{\mathit{DV}}$. 
As discussed above, the x-axis shows the normalised delta in percentage computed as $\Delta_{\mathit{DV}}= \frac{D-V}{V}*100$ and $\Delta_{\mathit{DN}}= \frac{D-N}{N}*100$.
The x-axis counts the number of specifications showing a given
normalised overhead value.

In Figure~\ref{fig:rq4-time}, we observe that the time overhead
$\Delta^t_{\mathit{DN}}$, is under \qty{30}{\percent} for 91 out of 112 specifications.
Similarly, the time overhead $\Delta^t_{DV}$ is under \qty{30}{\percent} for 81 specifications.
Similar values of $\Delta^t_{\mathit{DN}}$ and
$\Delta^t_{\mathit{DV}}$ for most of our specifications indicate that
the execution times for non-instrumented and vanilla-instrumented
programs are very close.
Thus, we deduce that most of the overhead comes from the instrumentation for the diagnosis rather than the instrumentation for the sole monitoring of the specification (i.e., the vanilla instrumentation points).
This is coherent with the fact that the diagnosis instrumentation
points include both the vanilla instrumentation points and all those program points that can affect them.
Finally, by inspecting the outliers where D takes more than \qty{200}{\percent} compared to N and V, we note that they are associated with traces with at least half million events.
For these specifications, the diagnostics requires instrumenting program points within loops executed numerous times that therefore generate a large number of events.
In Figure~\ref{fig:rq4-memory}, we observe for peak memory overhead similar trends as for time.
We see that both for $\Delta^m_{\mathit{DN}}$ and for
$\Delta^m_{\mathit{DV}}$ almost all of our specifications (\num{110}
out of \num{112}) have under \qty{20}{\percent} increase in memory consumption.
Only two specifications have an overhead over \qty{20}{\percent} and only one of those has an overhead over \qty{100}{\percent}.
\begin{Sbox}
\begin{minipage}{0.95\linewidth}
    {\bf RQ4 answer.} In the vast majority of our experiments, \icftlDiagnostics shows a limited overhead of less than \qty{30}{\percent} in time and less than \qty{20}{\percent} in peak memory.
    While not negligible, this overhead is still limited and, depending on
    the specific application, it can be out-weighted by the gained
    diagnosis information.
\end{minipage}
\end{Sbox}
\shadowbox{\TheSbox}

\subsection{Threats to validity}
To answer RQ1, we relied on a manually constructed ground truth of relevant program points for each specification, due to the absence of established diagnosis benchmarks in the literature. 
This manual process is susceptible to human error and subjective interpretation. 
To mitigate this threat, the first two authors independently annotated the ground truth and cross-validated their results, and discussed all discrepancies. 

Another potential threat comes from the manual definition of the specifications, which may limit the generalisability of our experiments to real-world applications. Since we, rather than the developers of the projects, wrote the specifications, they might not fully capture meaningful or representative behaviours of the systems under study. 
However, we based our specifications on available test cases and documentation, either by formalising constraints checked in assertions, or by covering code exercised during test execution.
This approach increases our confidence that the specifications are reasonably aligned with what a project developer might have written.

The measurement of memory usage and execution time---both for the diagnostic process (RQ3) and instrumentation overhead (RQ4)---is subject to variability, particularly in a language like Python, which lacks strict performance guarantees. 
To mitigate this threat, we used repeated experiments to average out the variability.
Furthermore, the results show consistently a low demand of time and computational resources, with few exceptions.
This gives confidence on the limited computational demands of \icftlDiagnostics and thus on our conclusions about its practical applicability.

     \section{Related work}
\label{sec:related-work}
Our work focuses on identifying the program points that contribute to how a variable obtains its value.
This goal intersects with several areas of software engineering, which include trace diagnostics, fault localisation (FL), causality analysis, and vulnerability analysis.

\paragraph{Trace diagnostics} 
Our approach belongs to the field of trace diagnostics, aiming to explain why a specification is violated by analysing execution traces.
A number of trace diagnostics techniques have been proposed for both source code-level and signal-level specifications.
Notably, \citet{dawes2019explaining} proposed a diagnostic method for CFTL (the predecessor of \icftl), which identified problematic code segments using path profiling.
In contrast, our approach focuses on identifying the program points that contribute to how a variable obtains its value, rather than focusing on path profiling.
Other approaches focus on signal-based properties.
\citet{boufaied2023trace} proposed the TD-SB-TemPsy approach for trace
diagnostics of SB-TemPsy-DSL properties, including a catalogue of 34 violation causes, each linked to a corresponding diagnosis.
Similarly, \citet{dou2018model} proposed TemPsy-Report, a model-driven approach that retrieved diagnostic information from a predefined list of violation types.
\citet{ferrere2015trace} formulated the diagnostics problem as identifying a minimal portion of the input signal that is sufficient to imply the violation of a specification.
\citet{nivckovic2020amt} implemented in the AMT 2.0 tool two diagnostic procedures: one that computed minimal explanations based on the method of \citet{ferrere2015trace}, and another that returned extended explanations using epoch diagnostics.
\citet{bartocci2019automatic} proposed the CPSDebug tool, which explains failures in Simulink/Stateflow models by generating a sequence of snapshots that highlight faulty behaviours.
\citet{basnet2020logical} presented a signal-level diagnostic technique that transforms time-domain signals into the frequency domain to identify frequency components relevant to a temporal logic formula, thereby enabling monitoring-safe compression.
\citet{claus2005periodic} modelled program behaviour by segmenting execution traces into phases and fitting low-degree polynomials with periodic coefficients, enabling the capture of recurring patterns.
While our work shares the goal of explaining specification violations, it differs in the nature of the artefact being diagnosed.
\icftlDiagnostics targets violations over the system’s source code, whereas the aforementioned approaches focus on signal traces or model-level behaviours.

\paragraph{Fault localisation} 
Our approach is related to FL, as both aim to identify program points that are relevant to explaining unexpected program behaviour.
FL typically focuses on locating the source of a failure in a program and ranking code statements based on their likelihood of being faulty.
Many FL techniques focus on ranking code statements by a suspicion score using passed and failed tests~\cite{wong2007effective}.
These techniques, such as the ones surveyed by~\citet{wong2016survey}, aim to identify the most likely faulty code segments by analysing the execution traces of successful and failed test cases.
Other approaches go further and also aim to create patches for program repair.
For instance, \citet{pearson2017evaluating} evaluated various automated program repair techniques that generate patches to fix faults.
Similarly, Louloudakis et al.~\cite{louloudakis2024fixcon} proposed Fix-Con, an automated approach for FL and repair in deep learning models.
From the failed tests, some techniques automatically isolate the cause-effect chain to pinpoint the root cause of failures.
Some notable examples include,~\citet{zeller2002isolating}'s work on isolating cause-effect chains and~\citet{cleve2005locating}'s work on locating causes of program failures.
\citet{liu2006failure} introduced the concept of failure proximity to identify similar fault locations, enhancing the precision of FL.
While our work shares the goal of identifying relevant program points, it extends beyond the typical objectives of FL, constructing a diagnosis that explains how a variable obtains its value, contributing to program comprehension, verification, and reasoning about correct behaviour.

\paragraph{Causality analysis}
Our approach shares with causality analysis the goal of identifying chains of events that lead to specific behaviours in the analysed system.
In the context of software, causality analysis focuses on determining how specific inputs, states, or events lead to particular outcomes, especially failures.
\citet{zeller2002isolating}’s Cause-Effect Chains looked at the sequence
of state changes leading to a failure, which helped developers understand the progression of the fault through the program’s execution.
\citet{cleve2005locating}'s Cause Transitions, instead, focused on identifying critical moments where the responsibility for the failure shifts from one variable to another, offering precise defect localisation. 
\citet{feng2024rocasrootcauseanalysis} contributed to causality analysis by providing a structured approach to identify the root causes of accidents in Autonomous Driving Systems (ADS). 
While our approach also inspects system behaviour during execution, causality analysis typically centres on identifying state changes that directly lead to failures or bugs.
\icftlDiagnostics focuses on identifying relevant program points that explain expressions in violated atoms, where an expression may not directly represent a fault or failure. 
Instead, it may reflect a combination of events that together lead to a specification violation. 

\paragraph{Vulnerability analysis}
Our approach shares with vulnerability analysis the goal of identifying critical program points that influence undesired outcomes.
Vulnerability analysis is an area in the domain of cybersecurity that aims to identify, evaluate, and address security weaknesses in systems, networks, and applications.
The goal is to discover vulnerabilities before they can be exploited by malicious actors.
\citet{THOME2018security} proposed an approach that assists security auditors in identifying and fixing common injection vulnerabilities in web applications. 
The main commonality with our diagnostics approach lies in the proposed “security slices”, which are reduced versions of program slices that contain only the essential information needed for security auditing. However, our approach is not limited to security contexts.
Joern~\cite{joernGitHub}, a static analysis platform that generates code property graphs, also supports vulnerability discovery by enabling expressive graph-based queries over code structure and semantics.
While our approach also uses graph-based representations to identify instrumentation points, it differs by combining static analysis with runtime monitoring.
\icftlDiagnostics captures concrete states that correspond to these
instrumentation points to construct a diagnosis, going beyond
vulnerability detection to explain how specific values evolve during
execution.

    \section{Conclusion and Future work}
\label{sec:concl-future-work}
In this paper, we have presented a novel approach for diagnosing state-based \icftl specifications.
Our approach relies on the diagnostics instrumentation that statically identifies all the instrumentation points relevant to a given expression in a specification.
After program execution, we introduced a new \iotatrace filtering algorithm that takes these instrumentation points and identifies all corresponding concrete states within the \iotatrace.
In this way, for each expression within a falsified atom of a specification, we provide the set of concrete states---i.e., statements executions---that have contributed to the computation of the expression value that falsified the specification.
We complement this analysis with the multiplicity metric, which measures the significance of each event as the number of distinct data flow paths linking it to the expression that caused the specification violation.
We have evaluated our approach on \num{10} different projects and \num{112} specifications, in terms of effectiveness, ability to reduce the complexity of understanding specifications, computational cost, and instrumentation overhead.
Empirically, \icftlDiagnostics identified relevant events with \SI{90}{\percent} precision and reduced code inspection by over \SI{90}{\percent}, while keeping diagnosis time under \SI{7}{\min} and instrumentation overhead below \SI{30}{\percent}.

Future research directions include extending our diagnostic framework to handle specifications that combine both time-based and state-based atoms. 
Such an integration would yield more comprehensive and informative diagnosis compared to getting a diagnosis for each atom in isolation.
Additionally, we plan to extend \icftlDiagnostics to support both source code and signal-based behaviours; a suitable language for expressing such hybrid specifications is SCSL~\cite{dawes2022scsl}.
This direction presents the challenge of designing diagnostic
techniques capable of seamlessly handling specifications that
integrate time-based, state-based, and signal-based constraints.

     \section*{Acknowledgments}
    The authors would like to thank Dr. Joshua Dawes for his guidance at the early stage of this research.
This research was funded in whole, or in part, by the Luxembourg National Research Fund (FNR), grant reference 17065054. For the purpose of open access, and in fulfilment of the obligations arising from the grant agreement, the author has applied a Creative Commons Attribution 4.0 International (CC BY 4.0) license to any Author Accepted Manuscript version arising from this submission.
    This project has received funding from the European Union’s Horizon Europe program under Grant Agreement No.\ 101148870 (ContTestCPS).
    The experiments presented in this paper were carried out using the
    HPC facilities of the University of
    Luxembourg~\cite{VCPKVO_HPCCT22} (see \url{https://hpc.uni.lu}).

    \bibliographystyle{IEEEtranN}

\end{document}